\documentclass[
    aps,prl,twocolumn,
    reprint,
    superscriptaddress,
    floatfix,
    amssymb,
    longbibliography
]{revtex4-1}

\usepackage{amsmath}
\usepackage{amsfonts}
\usepackage{graphicx}

\usepackage[utf8]{inputenc}

\usepackage{nicefrac}

\usepackage[normalem]{ulem}

\usepackage{xcolor}

\definecolor{linkColor}{rgb}{0,0.3,0.7}
\usepackage{hyperref}
\hypersetup{colorlinks=true,
            allcolors=linkColor,
            pdfborder={0 0 0},
            pdfencoding = auto
            }

\definecolor{darkblue}{RGB}{0,0,200}
\definecolor{darkred}{RGB}{200,0,0}

\def\etaStat{\eta_\mathrm{stat}}

\begin{document}

\title{Wavelength selection by interrupted coarsening in reaction--diffusion systems}

\author{Fridtjof Brauns} 
\email{fridtjof.brauns@lmu.de}
\author{Henrik Weyer}
\affiliation{Arnold Sommerfeld Center for Theoretical Physics and Center for NanoScience, Department of Physics, Ludwig-Maximilians-Universit\"at M\"unchen, Theresienstra\ss e 37, D-80333 M\"unchen, Germany}
\author{Jacob Halatek}
\affiliation{Biological Computation Group, Microsoft Research, Cambridge CB1 2FB, UK}
\author{Junghoon Yoon}
\author{Erwin Frey}
\email{frey@lmu.de}
\affiliation{Arnold Sommerfeld Center for Theoretical Physics and Center for NanoScience, Department of Physics, Ludwig-Maximilians-Universit\"at M\"unchen, Theresienstra\ss e 37, D-80333 M\"unchen, Germany}

	
\begin{abstract}
    Wavelength selection in reaction--diffusion systems can be understood as a coarsening process that is interrupted by counteracting processes at certain wavelengths. We first show that coarsening in mass-conserving systems is driven by self-amplifying mass transport between neighboring high-density domains. We derive a general coarsening criterion and show that coarsening is generically uninterrupted in two-component systems that conserve mass. The theory is then generalized to study interrupted coarsening and anti-coarsening due to weakly-broken mass conservation, providing a general path to analyze wavelength selection in pattern formation far from equilibrium.
\end{abstract}

\maketitle

To predict the wavelength of patterns in highly nonlinear systems is a critical open problem as wavelength selection is ubiquitous in a large range of non-equilibrium systems \cite{Liu.Goldenfeld1989, Glotzer.etal1995, Caussin.etal2014, Solon.etal2015, Murray.Sourjik2017, Chiou.etal2020, Gai.etal2020}.
While the amplitude equation formalism and weakly nonlinear analysis have been highly successful in the vicinity of onset \cite{Cross.Hohenberg1993}, these approaches are not informative for large amplitude patterns far away from onset. 
For one-component systems, a theory for wavelength selection based on a multiple-scale analysis has been developed~\cite{Politi.Misbah2004,Politi.Misbah2006}, but generalizations to multi-component systems have remained elusive.

In this Letter, we propose that wavelength selection in reaction--diffusion systems can be understood as a coarsening process that is interrupted and even reversed by counteracting processes at certain wavelengths.
Specifically, we study two-component systems and develop a theory for the mass-conserving case first where coarsening is uninterrupted. We then generalize this theory to account for source terms that break mass conservation and counteract the coarsening process.

While coarsening is well understood as minimization of the free energy for systems relaxing to thermal equilibrium (such as binary mixtures \cite{Wagner1961,Lifshitz.Slyozov1961}), this reasoning is generally not applicable for non-equilibrium systems such as most reaction--diffusion systems.
Two-component mass-conserving reaction--diffusion (MCRD) systems serve as paradigmatic models for intracellular pattern formation \cite{Otsuji.etal2007,Goryachev.Pokhilko2008,Altschuler.etal2008,Mori.etal2008, Jilkine.Edelstein-Keshet2011, Trong.etal2014,Chiou.etal2018}, and are used as phenomenological models for a wide range of systems including precipitation patterns~\cite{Scheel2009}, granular media~\cite{Aranson.Tsimring2008}, and braided polymers~\cite{Forte.etal2019}.
It has long been speculated that two-component MCRD systems generically exhibit uninterrupted coarsening~\cite{Ishihara.etal2007,Mori.etal2008,Chiou.etal2018,Morita.Ogawa2010}.
However, it has remained unclear whether coarsening always goes to completion in two-component MCRD systems, largely owing to a lack of insight into the underlying physical processes.

Here, we show that coarsening is driven by positive feedback in the competition for mass, derive a simple and quantitative description of coarsening dynamics, and explain why coarsening is generically uninterrupted in two-component MCRD systems.
As they are grounded in a phase-space analysis \cite{Brauns.etal2020d}, our results are independent of the specific mathematical form of the reaction kinetics.

Building on the insights into the coarsening process in the mass-conserving case, we elucidate and quantify the physical mechanisms underlying wavelength selection in the presence of weak source terms (weakly broken mass conservation). Coarsening arrests when mass competition is balanced by production and degradation. Moreover, domain splitting---owing to the destabilization of plateaus---reverses coarsening. Both are graphically understood by a generalization of the phase-space analysis.
Since our approach builds on studying the spatial redistribution of a nearly conserved quantity, we expect that it can be generalized beyond two-component reaction--diffusion systems; for instance, to systems with more components and to hydrodynamic models for active matter systems~\cite{Cates.etal2010, Caussin.etal2014, Cates.Tailleur2015, Liebchen.Levis2017, Chate2020}.

The general form of a reaction--diffusion system with two components, $u$ and $v$, can be written as
\begin{subequations} \label{eq:2CRD}
\begin{align}
	\partial_t u(x,t) &= D_u \nabla^2 u + f(u,v) + \varepsilon \, s_1(u,v), \\
	\partial_t v(x,t) &= D_v \nabla^2 v - f(u,v)+ \varepsilon \, s_2(u,v),
\end{align}
\end{subequations}
on a domain $\Omega$, with either no-flux or periodic boundary conditions \footnote{We discuss the dispersion relation of Eq.~\eqref{eq:2CRD} linearized around a homogeneous steady state in Fig.~S1 in the SM.}.
For specificity, we choose $D_u < D_v$ \footnote{Our findings immediately generalize to systems with density-independent cross-diffusion, see SM Sec.~1.2.}.
The reaction term $f$ describes conversion between $u$ and $v$ while the source terms $s_{1,2}$ with a (small) dimensionless source strength $\varepsilon$ break mass conservation.

\begin{figure}
	\includegraphics{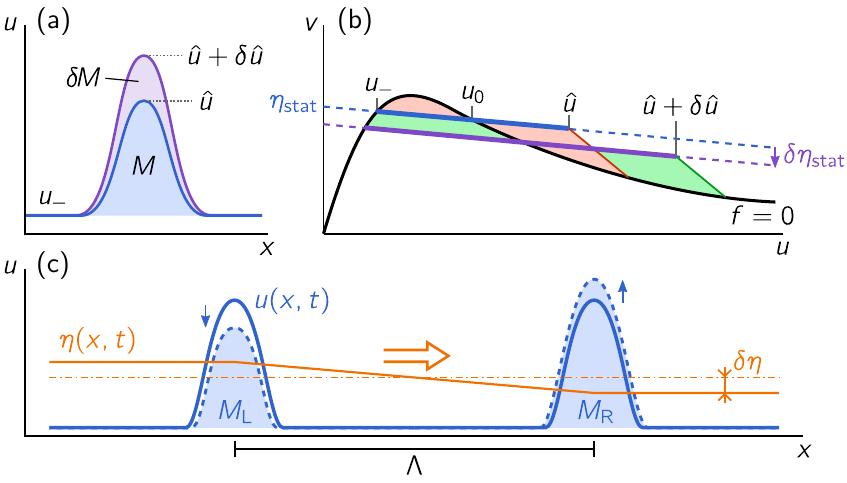}
	\caption{
	(a) Illustration of a stationary peak with peak mass $M$. Increasing the mass to $M + \delta M$ increases the peak amplitude to $\hat{u} + \delta \hat{u}$. 
	(b) Representation of the stationary peak in phase space (thick blue line), which is constrained to the FBS (dashed blue line). The FBS-offset $\etaStat(M)$ is determined by a balance of total reactive turnovers (areas shaded in red).
	For a peak with increased mass $M + \delta M$, and thus increased peak amplitude $\delta \hat{u}$, the FBS shifts downwards $\delta \etaStat$ until total turnover balance is restored (balance of green-shaded areas).
	(c) After a perturbation of two identical stationary peaks, the gradient in the mass-redistribution potential $\eta$ (orange line) drives mass-transport between the peaks (orange arrow) such that the larger (smaller) peak grows (shrinks) further (blue arrows).
	}
	\label{fig:1}
\end{figure}

\begin{figure*}[bt]
	\includegraphics{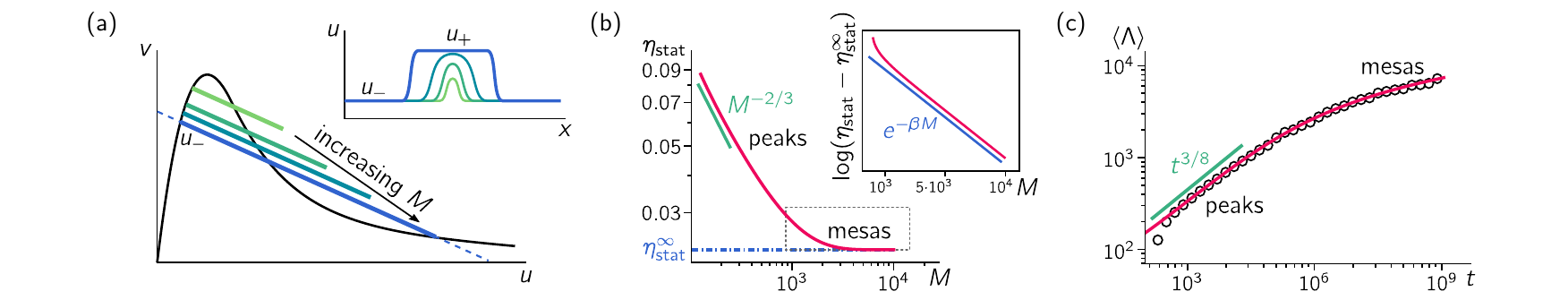}
	\caption{
	(a) Illustration of the peak to mesa transition as the total mass $M$ is increased.
	(b) The function $\etaStat(M)$ obtained by numerical continuation of the stationary solutions for the reaction kinetics $f_\mathrm{ex}$. Crossover from power law for peak patterns (amplitude not saturated) to exponential approach to $\etaStat^\infty$ for mesa patterns. 
	(c) Coarsening dynamics from finite element simulations for $f_\mathrm{ex}$ (black circles; mean peak distance averaged over four independent runs started from random initial conditions; parameters: $D_u = 1, D_v = 10^4, \bar{\rho} = 1.5$ and system size $|\Omega| = 2 \times 10^5$, periodic boundary conditions).
	The red line shows the analytic prediction based on $\sigma_\mathrm{D}^{}$ from $\etaStat(M)$, shown in (b), via Eq.~\eqref{eq:rate-diff-limited}. After an initial transient, power-law coarsening $\Lambda\sim t^{3/8}$ for peaks is observed, which flattens into logarithmic coarsening for mesas.
	}
	\label{fig:peak-and-mesa-coarsening}
\end{figure*}

Let us first analyze the mass-conserving case $\varepsilon = 0$. Then, the total density $\rho = u+v$ is conserved such that the average $\bar{\rho} = |\Omega|^{-1} \int_\Omega \mathrm dx \rho (x,t)$ remains constant.
The time evolution of $\rho$ is given by \cite{Otsuji.etal2007,Ishihara.etal2007,Brauns.etal2020d,Forte.etal2019}
\begin{equation} \label{eq:dtn}
 	\partial_t \rho(x,t) = D_v \nabla^2 \eta(x,t)
\end{equation}
with the \emph{mass-redistribution potential} defined by $\eta := v + (D_u/D_v) u$;
the corresponding dynamical equation for $\eta(x,t)$ is given in the SM~\cite{Note:SM}, Sec.~1.1.
For stationary patterns $[\tilde{u}(x),\tilde{v}(x)]$, the mass-redistribution potential must be spatially uniform, $\eta(x) = \etaStat$.
Based on this one can analyze two-component MCRD systems in the $(u,v)$ phase plane~\cite{Brauns.etal2020d}:
There, stationary patterns are constrained to a linear subspace, $v + (D_u/D_v) u = \etaStat$, called \emph{flux-balance subspace} (FBS); see Fig.~\ref{fig:1}b.
The intercept $\etaStat$ is determined by the balance of the spatially integrated reactive flows (\emph{total turnover balance}), corresponding (approximately) to a balance of areas (shaded in red in Fig.~\ref{fig:1}b) enclosed by the FBS and the \emph{reactive nullcline} ($f = 0$, NC).
The FBS-NC intersection points correspond to the plateau(s) and inflection point(s) of a stationary pattern.
Two types of patterns can be distinguished---mesas and peaks. The elementary mesa pattern is composed of two plateaus, connected by an interface (or ``kink''), while a peak forms when the maximum density does not saturate in a high-density plateau (Fig.~\ref{fig:1}a, compare Fig.~\ref{fig:peak-and-mesa-coarsening}a) \cite{Brauns.etal2020d}\footnote{Diffusive flux balance ensures that any stationary pattern can be dissected by inserting no-flux boundaries at its extrema.}.
We begin the analysis with peak patterns and then generalize the results to mesas.

\paragraph{A mass-competition instability drives coarsening.}
Coarsening requires the transport of mass between peaks.
Because mass transport is diffusive, it is fastest on the shortest length scales; hence, the dominant process is competition for mass between neighboring peaks (Fig.~1a).
Thus, as an elementary case, we study two peaks in a `box' with no-flux boundary conditions.
Consider a situation (``coarsening limit'') where the peaks are well separated, such that diffusive transport is limiting.
We can then approximate the peaks to be in (regional) quasi-steady state (QSS), such that $\eta = \etaStat(M)$ at a given peak with total mass $M$. This approximation is commonly applied in thin film theory \cite{Glasner.Witelski2003,Pismen.Pomeau2004} and Ostwald ripening \cite{Lifshitz.Slyozov1961,Wagner1961}.

Starting from two identical, stationary peaks, each with total mass $M_0$,
the dynamics of the mass difference between them ($M_\mathrm{R,L} = M_0 \pm \delta M$)---obtained by integration of Eq.~\eqref{eq:dtn} over a single peak---is determined by the $\eta$-gradients in the plateau between them (indicated by the orange arrow in Fig.~\ref{fig:1}c).
Using QSS at each peak separately, the mass-redistribution potential at the peaks is given by $\eta_\mathrm{R,L}^{} = \etaStat \pm (\partial_M^{}\etaStat|_{M_0}) \, \delta M$.
Between the peaks, $\eta$ obeys $\partial_x^2\eta=0$ because diffusive relaxation within the plateau is fast compared to the peak evolution (see SM Sec.~2 for details). Thus in 1D, the resulting gradient in $\eta$ is linear and determined by $\eta = \eta_\mathrm{R,L}^{}$ at the peak positions.
For a given peak separation $\Lambda$, this approximation determines the dynamics of mass redistribution
\begin{equation} \label{eq:rate-diff-limited}
    \partial_t \delta M \approx -\frac{2 D_v}{\Lambda} \left(\partial^{}_M \etaStat \big|_{M_0}\right) \delta M =: \sigma_\mathrm{D}\,\delta M.
\end{equation}
The subscript D denotes the diffusion-limited regime.
If the growth rate $\sigma_\mathrm{D}$ is positive, an instability driven by positive feedback in competition for mass results in coarsening.
Hence, the condition for uninterrupted coarsening reads
\begin{equation} \label{eq:coarsening-crit}
    \partial_M^{} \etaStat(M) < 0,
\end{equation}
i.e.\ that $\etaStat(M)$ is a strictly monotonically decreasing function for all stable stationary single-peak solutions.
This recovers a previous, mathematically derived coarsening condition~\cite{Ishihara.etal2007,Otsuji.etal2007}.
Importantly, the analysis presented here gives insight into the underlying physical mechanism and shows that not only the criterion for coarsening, but the entire \emph{temporal evolution} of coarsening is determined by $\partial_M^{} \etaStat$ via Eq.~\eqref{eq:rate-diff-limited} \footnote{For mesa patterns in 1D, $\partial_M\etaStat$ must be calculated for the high- and low-density plateaus separately (see SM Sec.~5).}.
We learn that the functional dependence of the \emph{mass-redistribution potential} on the peak mass, $\etaStat(M)$, plays a role analogous to the functional dependence of the chemical potential on the droplet size that drives Ostwald ripening or to the film height in dependence of droplet size that drives coarsening of unstable thin films \cite{Glasner.Witelski2003,Pismen.Pomeau2004}.

\paragraph{Generic coarsening laws for mass-conserving systems.}
To show that coarsening is uninterrupted, we need to show that the criterion Eq.~\eqref{eq:coarsening-crit} holds, and continues to hold as small peaks disappear causing the mass of the remaining peaks to increase. For an intuitive argument, consider a single stationary peak with mass $M$ (see Fig.~\ref{fig:1}a) and its representation in phase space, the blue line in Fig.~\ref{fig:1}b.
Add an amount $\delta M$ of mass and hold $\etaStat$ fixed for the moment (for the sake of argument).
Fixing $\etaStat$ also fixes the plateau $u_-$.
Therefore, the additional mass will increase the peak amplitude $\hat{u}$ (Fig.~\ref{fig:1}b), causing the reactive turnover to the right of $u_0$ to increase.
The resulting imbalance of total turnover entails a net reactive flow that shifts the flux-balance subspace downwards, i.e.\ lowers $\etaStat$, to restore total turnover balance.
We conclude that, $\etaStat(M)$ is generically a monotonically decreasing function.
(More rigorous arguments are given in SM Secs.~4 and~5).

Let us now turn to the dynamic coarsening laws.
As an example, consider $f_\mathrm{ex} \,{=}\, (1 \,{+}\, u)v \,{-}\, u/(1 \,{+}\, u)$, where the first and second terms may, for instance, describe protein recruitment and first-order enzymatic detachment, respectively.
A simple scaling argument~
\footnote{For a back of the envelope calculation we use density at the pattern inflection point $u_0$ and the interface width approximation $\ell_\mathrm{int} \,{\approx}\, \pi /q_\mathrm{max}|_{u_0, \etaStat}$ \cite{Brauns.etal2020d}, and $\hat{u} \approx 2 u_0$ to estimate the peak mass $M \,{=}\, \int \mathrm{d}x [\tilde{\rho}(x) \,{-}\, \rho_-] \,{\approx}\, \hat{u} \cdot \ell_\mathrm{int} \,{\approx}\, u_0 \cdot 2 \pi /q_\mathrm{max}|_{\etaStat} \,{\approx}\, 2\pi u_0 \sqrt{D_u/f_u|_{\etaStat}}$. For the example $f_\mathrm{ex}$, one finds $f_u \,{\approx}\, \etaStat$ and $u_0^{} \,{\approx}\, \etaStat^{-1}$, and thus $M \,{\sim}\, \etaStat^{-3/2} \,{=}\, \etaStat^{-1/\alpha}$, for sufficiently large $M$ and $D_v \,{\gg}\, D_u$.
Other reaction terms, e.g.\ with different nonlinearities in the recruitment term, yields other exponents $\alpha$.
} 
yields a power-law relation $\etaStat(M) \,{\sim}\, M^{-\alpha}$,
where the exponent depends on the specific reaction kinetics ($\alpha \,{=}\, \nicefrac23$ for the example above); see Fig.~\ref{fig:peak-and-mesa-coarsening}b.
In a large system containing multiple peaks, the average peak separation $\langle\Lambda\rangle$ is linked to the characteristic peak mass by $\langle M\rangle \,{=}\, (\bar{\rho} \,{-}\, \rho_-) \langle\Lambda\rangle$, where $\rho_-$ is the total density in the low density plateau between the peaks, and $\langle\cdot\rangle$ denotes an average over the entire system.
As peaks collapse, with a typical time given by the inverse growth rate of the mass-competition instability $t \,{\sim}\, \sigma_\mathrm{D}^{-1}$, the average peak separation $\langle\Lambda\rangle$ will increase. 
Combining  $\sigma_\mathrm{D} \,{\sim}\, {-}\langle\partial_M^{}\etaStat\rangle/\langle\Lambda\rangle$ with $\langle \partial_M^{} \etaStat \rangle \,{\sim}\, \langle M \rangle^{-\alpha-1} \,{\sim}\, (\bar{\rho}\langle\Lambda\rangle)^{-\alpha-1}$ yields power-law coarsening with $\langle\Lambda\rangle(t) \,{\sim}\, t^{1/(2 + \alpha)}$; see Fig.~\ref{fig:peak-and-mesa-coarsening}c and Fig.~S4.
Moreover, using appropriate scaling amplitudes, the coarsening trajectories for different average masses $\bar{\rho}$ can be collapsed onto a single master curve obtained from $\partial_M \etaStat$ (see SM Sec.~3).
Power-law coarsening in 1D has previously been found for peak-like droplets formed during the dewetting of thin liquid films \cite{Glasner.Witelski2003}.

As peaks collapse, those remaining grow in mass and height.
When the density at the peak maximum saturates in a high-density plateau (corresponding to a FBS-NC intersection point in phase space), a mesa pattern starts to form (Fig.~\ref{fig:peak-and-mesa-coarsening}a, Fig.~S3) \footnote{If $u,v$ describe (shifted) concentrations ($u,v$ bounded from below) and $D_u/D_v$ is finite, mesas inevitably form at high densities (cf.\ Fig.~\ref{fig:peak-and-mesa-coarsening}a). Arbitrarily large peaks form if no third FBS-NC intersection point exists; as, for example, in ``Model II'' in Ref.~\cite{Otsuji.etal2007}.
The peak-to-mesa transition has previously been observed for unstable thin films subject to gravity \cite{Gratton.Witelski2008}.
}.
\phantom{\cite{Gratton.Witelski2008}} 
For such mesas, somewhat more subtle arguments show that $\etaStat(M)$ remains a monotonically decreasing function (see SM Sec.~5).
In essence, changing $M$ shifts the interface positions and thus changes the width of a mesa's plateau.
As the density profile approaches the limiting plateaus $u_\pm(\etaStat^\infty)$ through exponential tails, $\etaStat(M)$ approaches $\etaStat^\infty$ exponentially slowly (see inset in Fig.~2b) where we define $\etaStat^\infty$ as the limit of $\etaStat$ for the stationary pattern on an infinite domain (see SM Sec.~5.1).
Using the same scaling arguments as for peaks, one obtains a logarithmic coarsening law for all mesa patterns, as in the one-dimensional Cahn--Hilliard model~\cite{Langer1971}.
For the concrete example $f_\mathrm{ex}$, we find excellent agreement between finite-element simulations and $\langle\Lambda\rangle(t)$ obtained from $\etaStat(M)$ by these scaling arguments (see Fig.~\ref{fig:peak-and-mesa-coarsening}c).
Based on the physical insights presented above, a generalization to more than one spatial dimension is straightforward. For mesa-like droplets with radius $R$ one finds $\etaStat-\etaStat^\infty \sim R^{-1}$ which yields power law coarsening with the universal exponent $\nicefrac{1}{3}$ (see SM Sec.~5.4). For peak-like droplets, we expect system-dependent exponents as in 1D.

\begin{figure*}[bt]
	\includegraphics{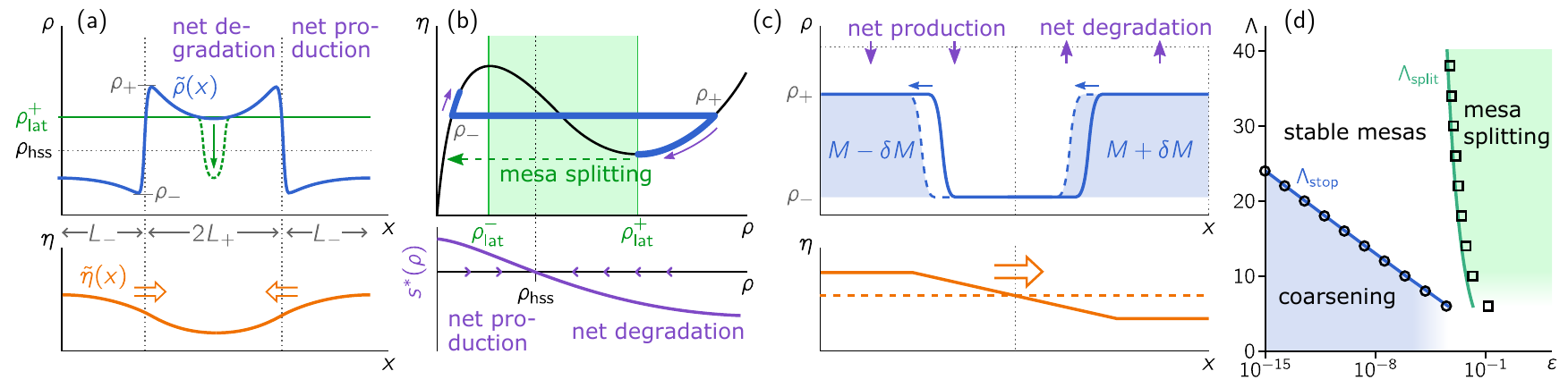}
	\caption{
	Wavelength selection by weakly broken mass conservation.
	(a,b)~Mesa splitting: (a) real space profiles of $\tilde{\rho}(x)$ and $\tilde{\eta}(x)$, (b)~phase space in $(\rho,\eta)$ coordinates,
	with the source term in local equilibrium approximation plotted below. The green shaded area indicates the region of lateral instability.
	(c)~Interrupted coarsening due to a balance of production, degradation, and mass redistribution between neighboring mesas.
	(d)~Regimes separated by interrupted coarsening (squares) and mesa splitting (circles) as well as analytic approximations (blue, green lines) for large $\Lambda$ and small $\varepsilon$. While in the coarsening regime (blue) stationary patterns are unstable, no stationary patterns exist in the mesa-splitting regime (green). In the regime of small $\Lambda$ and large $\varepsilon$, corrections become large and the approximations do not hold (see SM Sec.~7).
	Parameters: $D_u = 0.1, D_v = 1, p = 2$.
	}
	\label{fig:weakly-broken}
\end{figure*}

\paragraph{The limit of large $D_v$.}
For $D_v \,{\rightarrow}\, \infty$, mass redistribution by $v$-diffusion becomes instantaneous, such that the reactive conversion between $u$ and $v$, which drives the growth/shrinking of mesas or peaks, becomes limiting.
In this \emph{reaction-limited} case, we find $\sigma_\mathrm{R} \,{\approx}\, (\partial_M^{} \etaStat) \, \ell_\mathrm{int} \langle f_v \rangle_\mathrm{int}$, where $\ell_\mathrm{int}$ is the interface width and $\langle \cdot \rangle_\mathrm{int}$ denotes the average over the interface region (see SM Sec.~6 for details and numerical verification).
Comparing with Eq.~\eqref{eq:rate-diff-limited} shows that the coarsening criterion Eq.~\eqref{eq:coarsening-crit} holds in both regimes, and the crossover from diffusion- to reaction-limited coarsening occurs at $D_v/\Lambda \,{\approx}\, \ell_\mathrm{int} \langle f_v \rangle_\mathrm{int}$.

\paragraph{Weakly broken mass conservation.}
With an understanding for the coarsening dynamics in the strictly mass-conserving system, we now consider the effect of slow production and degradation for $0 < \varepsilon \ll 1$.
We will see that these additional processes interrupt coarsening \cite{Liu.Goldenfeld1989,Glotzer.etal1995,Kolokolnikov.etal2006, Cates.etal2010} and can reverse it by inducing peak/mesa splitting \cite{Kolokolnikov.etal2007, Murray.Sourjik2017, Gai.etal2020}, thus selecting a range of stable pattern wavelengths.
In the presence of source terms, the time evolution of the total density $\rho$ is governed by
\begin{equation} \label{eq:dtn-weakly-broken}
 	\partial_t \rho = D_v \partial_x^2 \eta + \varepsilon \, s(u,v),
\end{equation}
with the total source $s := s_1 + s_2$.
Hence, the average mass $\langle\rho\rangle$ is no longer a control parameter but a time-dependent variable that is determined indirectly by a balance of production and degradation (in short:
\emph{source balance}). 
In phase space, there are now two reactive nullclines, one each for $u$ and $v$, which both converge to $f \,{=}\, 0$ for $\varepsilon \,{\rightarrow}\, 0$.
Their intersection point(s) determine(s) the homogeneous steady state (HSS) $\rho_\mathrm{hss}$ that balances the total source term.

In the following, we restrict ourselves to mesa patterns. To lowest order in $\varepsilon$, source balance determines the `half lengths' $L_\pm$ of the upper and lower plateaus (see SM Sec.~7).
Along the plateaus, the spatial gradients induced by slow production--degradation ($\varepsilon$ small) are shallow, such that the dynamics is (approximately) slaved to the nullcline $f \,{=}\, 0$ (see Fig.~\ref{fig:weakly-broken}b). This justifies a local equilibrium approximation $s(u,v) \,{\approx}\, s[u^*(\rho), v^*(\rho)] \,{\equiv}\, s^*(\rho)$ in Eq.~\eqref{eq:dtn-weakly-broken}, where the local equilibria are defined by $f(u^*,v^*) \,{=}\, 0$ and $u^* \,{+}\, v^* \,{=}\, \rho$.
On the short scale of the interface width, the weak source term is negligible and each interface constrained to a flux-balance subspace.
We are now in a position to generalize the phase-space analysis introduced in Ref.~\cite{Brauns.etal2020d} and analyze interrupted coarsening and mesa splitting.

(\textit{i}) \textit{Peak/mesa splitting.}
Consider the fully coarsened state for $\varepsilon \,{=}\, 0$ and add a small source term such that  $s^*(\rho_+) \,{<}\, 0$ and $s^*(\rho_-) \,{>}\, 0$ (i.e.\ $\rho_- \,{<}\, \rho_\mathrm{hss} \,{<}\, \rho_+$, see Fig.~\ref{fig:weakly-broken}b) 
\footnote{In steady state, net degradation in high-density regions ($\rho > \rho_\mathrm{hss}$) and net production in low-density regions ($\rho < \rho_\mathrm{hss}$) must balance.
}. 
The upper plateau is depressed by net degradation and is refilled by inflow from the interfaces that connect to the lower plateau where net production prevails.
The longer the plateaus (and the larger $\varepsilon$), the more they curve towards $\rho_\mathrm{hss}$.
Since $\rho_- \,{<}\, \rho_\mathrm{hss} \,{<}\, \rho_+$, $\rho(x)$ will eventually enter the interval of lateral instability $[\rho_\mathrm{lat}^-,\rho_\mathrm{lat}^+]$ (where $\partial_\rho \eta^* \,{<}\, 0$), triggering a nucleation event that results the splitting of the mesa (see Fig.~\ref{fig:weakly-broken}a and Movie~2).
A simple approximation for the threshold wavelength $\Lambda_\mathrm{split}(\varepsilon)$ where this happens is derived in the SM, Sec.~7.1. Comparison with numerical simulations shows excellent agreement (see Fig.~\ref{fig:weakly-broken}d).

(\textit{ii}) \textit{Interrupted coarsening.}
Intuitively, production and degradation can counteract the mass-competition instability.
To determine the corresponding length scale $\Lambda_\mathrm{stop}$ where coarsening arrests, we consider the stability of two neighboring, symmetric mesas.
A perturbation that moves a small amount of mass from one mesa to the other (Fig.~\ref{fig:weakly-broken}c) has two effects:
First, it shifts the mass-redistribution potential at the interfaces, leading to mass transport that further amplifies the perturbation with rate $\sigma_\mathrm{D}^{}(\Lambda) \delta M$ as in the strictly mass-conserving situation; cf.\ Eq.~\eqref{eq:rate-diff-limited}.
Second, the changed lengths $\delta L \,{=}\, \delta M / (\rho_+ \,{-}\, \rho_-)$ of the two mesas result in net production (degradation) in the shorter (longer) mesa with rate $\varepsilon |s^*(\rho_\mathrm{outer})| \delta L$ (indicated by the purple arrows in Fig.~\ref{fig:weakly-broken}c).
Here $\rho_\mathrm{outer}$ denotes the total density of the outer plateau (the inner plateau shifts as a whole and does not change in length, see Fig.~\ref{fig:weakly-broken}b).
Together, the balance of both processes determines $\Lambda_\mathrm{stop}$ (see SM Sec.~7.2 for details)
\begin{equation} \label{eq:interrupted}
    \sigma_\mathrm{D}^{}(\Lambda_\mathrm{stop}) \approx \varepsilon \frac{|s^*(\rho_\mathrm{outer})|}{\rho_+-\rho_-}.
\end{equation}
As a concrete example, we apply Eq.~\eqref{eq:interrupted} to the ``Brusselator'' model \cite{Prigogine.Lefever1968} ($f \,{=}\, u^2 v \,{-}\, u$, $s \,{=}\, p \,{-}\, u$), and find excellent agreement with numerics (Fig.~\ref{fig:weakly-broken}c).
Notably, the simple estimate given by Eq.~\eqref{eq:interrupted} generalizes a previous, mathematically obtained results~\cite{Kolokolnikov.etal2006,McKay.Kolokolnikov2012}.

Our analysis shows that the mechanisms underlying mesa splitting and interrupted coarsening are distinct.
Notably, the length scale where coarsening stops is much smaller than the length scale where mesas/peaks split (see Fig.~\ref{fig:weakly-broken}d).
This implies that there are stable periodic patterns for a large, continuous range of wavelengths
(multistability), as was shown previously for the `Brusselator'
\cite{Prigogine.Lefever1968, Kolokolnikov.etal2006, Kolokolnikov.etal2007}. 
Similarly, multistability of wavelengths was recently found in a hydrodynamic model for flocking~\cite{Caussin.etal2014}. 
Interestingly, a unique length scale is selected once noise is accounted for~\cite{Solon.etal2015}.
Noise-driven wavelength selection was also observed in an ``active Model~B''~\cite{Tjhung.etal2018}.
It would be interesting to study whether this phenomenon is also found in reaction--diffusion systems.

Another interesting open problem are systems with cross diffusion and density-dependent diffusion coefficients (see e.g.\ Refs.~\cite{Vanag.Epstein2009,Rossi.etal2011,Giri.etal2020,Giunta.etal2020}).
We also expect that our approach can be generalized to systems with more than two components, higher spatial dimensions and also beyond reaction--diffusion systems.
In particular, conserved densities (particle numbers) are a generic feature of many active matter systems in which coarsening and length-scale selection (``micro-phase separation'') are of growing interest \cite{Cates.etal2010, Liu.etal2013, Wittkowski.etal2014, Caussin.etal2014, Cates.Tailleur2015, Gonnella.etal2015, Sabrina.etal2015, Liebchen.Levis2017, Tjhung.etal2018, Liu.etal2019, Curatolo.etal2019, Li.Cates2020, Chate2020}.

\begin{acknowledgments}
This work was funded by the Deutsche Forschungsgemeinschaft (DFG, German Research Foundation) through the Collaborative Research Center (SFB) 1032 -- Project-ID 201269156 -- and the Excellence Cluster ORIGINS under Germany’s Excellence Strategy -- EXC-2094 -- 390783311.

FB and HW contributed equally to this work.
\end{acknowledgments}


\begin{thebibliography}{71}%
\makeatletter
\providecommand \@ifxundefined [1]{%
 \@ifx{#1\undefined}
}%
\providecommand \@ifnum [1]{%
 \ifnum #1\expandafter \@firstoftwo
 \else \expandafter \@secondoftwo
 \fi
}%
\providecommand \@ifx [1]{%
 \ifx #1\expandafter \@firstoftwo
 \else \expandafter \@secondoftwo
 \fi
}%
\providecommand \natexlab [1]{#1}%
\providecommand \enquote  [1]{``#1''}%
\providecommand \bibnamefont  [1]{#1}%
\providecommand \bibfnamefont [1]{#1}%
\providecommand \citenamefont [1]{#1}%
\providecommand \href@noop [0]{\@secondoftwo}%
\providecommand \href [0]{\begingroup \@sanitize@url \@href}%
\providecommand \@href[1]{\@@startlink{#1}\@@href}%
\providecommand \@@href[1]{\endgroup#1\@@endlink}%
\providecommand \@sanitize@url [0]{\catcode `\\12\catcode `\$12\catcode
  `\&12\catcode `\#12\catcode `\^12\catcode `\_12\catcode `\%12\relax}%
\providecommand \@@startlink[1]{}%
\providecommand \@@endlink[0]{}%
\providecommand \url  [0]{\begingroup\@sanitize@url \@url }%
\providecommand \@url [1]{\endgroup\@href {#1}{\urlprefix }}%
\providecommand \urlprefix  [0]{URL }%
\providecommand \Eprint [0]{\href }%
\providecommand \doibase [0]{http://dx.doi.org/}%
\providecommand \selectlanguage [0]{\@gobble}%
\providecommand \bibinfo  [0]{\@secondoftwo}%
\providecommand \bibfield  [0]{\@secondoftwo}%
\providecommand \translation [1]{[#1]}%
\providecommand \BibitemOpen [0]{}%
\providecommand \bibitemStop [0]{}%
\providecommand \bibitemNoStop [0]{.\EOS\space}%
\providecommand \EOS [0]{\spacefactor3000\relax}%
\providecommand \BibitemShut  [1]{\csname bibitem#1\endcsname}%
\let\auto@bib@innerbib\@empty
\bibitem [{\citenamefont {Liu}\ and\ \citenamefont
  {Goldenfeld}(1989)}]{Liu.Goldenfeld1989}%
  \BibitemOpen
  \bibfield  {author} {\bibinfo {author} {\bibfnamefont {Fong}\ \bibnamefont
  {Liu}}\ and\ \bibinfo {author} {\bibfnamefont {Nigel}\ \bibnamefont
  {Goldenfeld}},\ }\bibfield  {title} {\enquote {\bibinfo {title} {Dynamics of
  phase separation in block copolymer melts},}\ }\href {\doibase
  10.1103/PhysRevA.39.4805} {\bibfield  {journal} {\bibinfo  {journal}
  {Physical Review A}\ }\textbf {\bibinfo {volume} {39}},\ \bibinfo {pages}
  {4805--4810} (\bibinfo {year} {1989})}\BibitemShut {NoStop}%
\bibitem [{\citenamefont {Glotzer}\ \emph {et~al.}(1995)\citenamefont
  {Glotzer}, \citenamefont {Di~Marzio},\ and\ \citenamefont
  {Muthukumar}}]{Glotzer.etal1995}%
  \BibitemOpen
  \bibfield  {author} {\bibinfo {author} {\bibfnamefont {Sharon~C.}\
  \bibnamefont {Glotzer}}, \bibinfo {author} {\bibfnamefont {Edmund~A.}\
  \bibnamefont {Di~Marzio}}, \ and\ \bibinfo {author} {\bibfnamefont
  {M.}~\bibnamefont {Muthukumar}},\ }\bibfield  {title} {\enquote {\bibinfo
  {title} {Reaction-{{Controlled Morphology}} of {{Phase}}-{{Separating
  Mixtures}}},}\ }\href {\doibase 10.1103/PhysRevLett.74.2034} {\bibfield
  {journal} {\bibinfo  {journal} {Physical Review Letters}\ }\textbf {\bibinfo
  {volume} {74}},\ \bibinfo {pages} {2034--2037} (\bibinfo {year}
  {1995})}\BibitemShut {NoStop}%
\bibitem [{\citenamefont {Caussin}\ \emph {et~al.}(2014)\citenamefont
  {Caussin}, \citenamefont {Solon}, \citenamefont {Peshkov}, \citenamefont
  {Chat{\'e}}, \citenamefont {Dauxois}, \citenamefont {Tailleur}, \citenamefont
  {Vitelli},\ and\ \citenamefont {Bartolo}}]{Caussin.etal2014}%
  \BibitemOpen
  \bibfield  {author} {\bibinfo {author} {\bibfnamefont {Jean-Baptiste}\
  \bibnamefont {Caussin}}, \bibinfo {author} {\bibfnamefont {Alexandre}\
  \bibnamefont {Solon}}, \bibinfo {author} {\bibfnamefont {Anton}\ \bibnamefont
  {Peshkov}}, \bibinfo {author} {\bibfnamefont {Hugues}\ \bibnamefont
  {Chat{\'e}}}, \bibinfo {author} {\bibfnamefont {Thierry}\ \bibnamefont
  {Dauxois}}, \bibinfo {author} {\bibfnamefont {Julien}\ \bibnamefont
  {Tailleur}}, \bibinfo {author} {\bibfnamefont {Vincenzo}\ \bibnamefont
  {Vitelli}}, \ and\ \bibinfo {author} {\bibfnamefont {Denis}\ \bibnamefont
  {Bartolo}},\ }\bibfield  {title} {\enquote {\bibinfo {title} {Emergent
  {{Spatial Structures}} in {{Flocking Models}}: {{A Dynamical System
  Insight}}},}\ }\href {\doibase 10.1103/PhysRevLett.112.148102} {\bibfield
  {journal} {\bibinfo  {journal} {Physical Review Letters}\ }\textbf {\bibinfo
  {volume} {112}},\ \bibinfo {pages} {148102} (\bibinfo {year}
  {2014})}\BibitemShut {NoStop}%
\bibitem [{\citenamefont {Solon}\ \emph {et~al.}(2015)\citenamefont {Solon},
  \citenamefont {Chat{\'e}},\ and\ \citenamefont {Tailleur}}]{Solon.etal2015}%
  \BibitemOpen
  \bibfield  {author} {\bibinfo {author} {\bibfnamefont {Alexandre~P.}\
  \bibnamefont {Solon}}, \bibinfo {author} {\bibfnamefont {Hugues}\
  \bibnamefont {Chat{\'e}}}, \ and\ \bibinfo {author} {\bibfnamefont {Julien}\
  \bibnamefont {Tailleur}},\ }\bibfield  {title} {\enquote {\bibinfo {title}
  {From {{Phase}} to {{Microphase Separation}} in {{Flocking Models}}: {{The
  Essential Role}} of {{Nonequilibrium Fluctuations}}},}\ }\href {\doibase
  10.1103/PhysRevLett.114.068101} {\bibfield  {journal} {\bibinfo  {journal}
  {Physical Review Letters}\ }\textbf {\bibinfo {volume} {114}},\ \bibinfo
  {pages} {068101} (\bibinfo {year} {2015})}\BibitemShut {NoStop}%
\bibitem [{\citenamefont {Murray}\ and\ \citenamefont
  {Sourjik}(2017)}]{Murray.Sourjik2017}%
  \BibitemOpen
  \bibfield  {author} {\bibinfo {author} {\bibfnamefont {Se{\'a}n~M.}\
  \bibnamefont {Murray}}\ and\ \bibinfo {author} {\bibfnamefont {Victor}\
  \bibnamefont {Sourjik}},\ }\bibfield  {title} {\enquote {\bibinfo {title}
  {Self-organization and positioning of bacterial protein clusters},}\ }\href
  {\doibase 10.1038/nphys4155} {\bibfield  {journal} {\bibinfo  {journal}
  {Nature Physics}\ }\textbf {\bibinfo {volume} {13}},\ \bibinfo {pages}
  {1006--1013} (\bibinfo {year} {2017})}\BibitemShut {NoStop}%
\bibitem [{\citenamefont {Chiou}\ \emph {et~al.}(2020)\citenamefont {Chiou},
  \citenamefont {Moran},\ and\ \citenamefont {Lew}}]{Chiou.etal2020}%
  \BibitemOpen
  \bibfield  {author} {\bibinfo {author} {\bibfnamefont {Jian-geng}\
  \bibnamefont {Chiou}}, \bibinfo {author} {\bibfnamefont {Kyle~D.}\
  \bibnamefont {Moran}}, \ and\ \bibinfo {author} {\bibfnamefont {Daniel~J.}\
  \bibnamefont {Lew}},\ }\bibfield  {title} {\enquote {\bibinfo {title} {How
  cells determine the number of polarity sites},}\ }\href {\doibase
  10.1101/2020.05.21.109520} {\bibfield  {journal} {\bibinfo  {journal}
  {bioRxiv}\ ,\ \bibinfo {pages} {doi:10.1101/2020.05.21.109520}} (\bibinfo
  {year} {2020})}\BibitemShut {NoStop}%
\bibitem [{\citenamefont {Gai}\ \emph {et~al.}(2020)\citenamefont {Gai},
  \citenamefont {Iron},\ and\ \citenamefont {Kolokolnikov}}]{Gai.etal2020}%
  \BibitemOpen
  \bibfield  {author} {\bibinfo {author} {\bibfnamefont {Chunyi}\ \bibnamefont
  {Gai}}, \bibinfo {author} {\bibfnamefont {David}\ \bibnamefont {Iron}}, \
  and\ \bibinfo {author} {\bibfnamefont {Theodore}\ \bibnamefont
  {Kolokolnikov}},\ }\bibfield  {title} {\enquote {\bibinfo {title} {Localized
  outbreaks in an {{S}}-{{I}}-{{R}} model with diffusion},}\ }\href {\doibase
  10.1007/s00285-020-01466-1} {\bibfield  {journal} {\bibinfo  {journal}
  {Journal of Mathematical Biology}\ } (\bibinfo {year} {2020}),\
  10.1007/s00285-020-01466-1}\BibitemShut {NoStop}%
\bibitem [{\citenamefont {Cross}\ and\ \citenamefont
  {Hohenberg}(1993)}]{Cross.Hohenberg1993}%
  \BibitemOpen
  \bibfield  {author} {\bibinfo {author} {\bibfnamefont {M.~C.}\ \bibnamefont
  {Cross}}\ and\ \bibinfo {author} {\bibfnamefont {P.~C.}\ \bibnamefont
  {Hohenberg}},\ }\bibfield  {title} {\enquote {\bibinfo {title} {Pattern
  formation outside of equilibrium},}\ }\href {\doibase
  10.1103/RevModPhys.65.851} {\bibfield  {journal} {\bibinfo  {journal}
  {Reviews of Modern Physics}\ }\textbf {\bibinfo {volume} {65}},\ \bibinfo
  {pages} {851--1112} (\bibinfo {year} {1993})}\BibitemShut {NoStop}%
\bibitem [{\citenamefont {Politi}\ and\ \citenamefont
  {Misbah}(2004)}]{Politi.Misbah2004}%
  \BibitemOpen
  \bibfield  {author} {\bibinfo {author} {\bibfnamefont {Paolo}\ \bibnamefont
  {Politi}}\ and\ \bibinfo {author} {\bibfnamefont {Chaouqi}\ \bibnamefont
  {Misbah}},\ }\bibfield  {title} {\enquote {\bibinfo {title} {When {{Does
  Coarsening Occur}} in the {{Dynamics}} of {{One}}-{{Dimensional Fronts}}?}}\
  }\href {\doibase 10.1103/PhysRevLett.92.090601} {\bibfield  {journal}
  {\bibinfo  {journal} {Physical Review Letters}\ }\textbf {\bibinfo {volume}
  {92}},\ \bibinfo {pages} {090601} (\bibinfo {year} {2004})}\BibitemShut
  {NoStop}%
\bibitem [{\citenamefont {Politi}\ and\ \citenamefont
  {Misbah}(2006)}]{Politi.Misbah2006}%
  \BibitemOpen
  \bibfield  {author} {\bibinfo {author} {\bibfnamefont {Paolo}\ \bibnamefont
  {Politi}}\ and\ \bibinfo {author} {\bibfnamefont {Chaouqi}\ \bibnamefont
  {Misbah}},\ }\bibfield  {title} {\enquote {\bibinfo {title} {Nonlinear
  dynamics in one dimension: {{A}} criterion for coarsening and its temporal
  law},}\ }\href {\doibase 10.1103/PhysRevE.73.036133} {\bibfield  {journal}
  {\bibinfo  {journal} {Physical Review E}\ }\textbf {\bibinfo {volume} {73}},\
  \bibinfo {pages} {036133} (\bibinfo {year} {2006})}\BibitemShut {NoStop}%
\bibitem [{\citenamefont {Wagner}(1961)}]{Wagner1961}%
  \BibitemOpen
  \bibfield  {author} {\bibinfo {author} {\bibfnamefont {C.}~\bibnamefont
  {Wagner}},\ }\bibfield  {title} {\enquote {\bibinfo {title} {{Theorie der
  Alterung von Niederschl\"agen durch Uml\"osen (Ostwald-Reifung)}},}\ }\href
  {\doibase 10.1002/bbpc.19610650704} {\bibfield  {journal} {\bibinfo
  {journal} {Zeitschrift f\"ur Elektrochemie}\ }\textbf {\bibinfo {volume}
  {65}},\ \bibinfo {pages} {581--591} (\bibinfo {year} {1961})}\BibitemShut
  {NoStop}%
\bibitem [{\citenamefont {Lifshitz}\ and\ \citenamefont
  {Slyozov}(1961)}]{Lifshitz.Slyozov1961}%
  \BibitemOpen
  \bibfield  {author} {\bibinfo {author} {\bibfnamefont {I.M.}\ \bibnamefont
  {Lifshitz}}\ and\ \bibinfo {author} {\bibfnamefont {V.V.}\ \bibnamefont
  {Slyozov}},\ }\bibfield  {title} {\enquote {\bibinfo {title} {The kinetics of
  precipitation from supersaturated solid solutions},}\ }\href {\doibase
  10.1016/0022-3697(61)90054-3} {\bibfield  {journal} {\bibinfo  {journal}
  {Journal of Physics and Chemistry of Solids}\ }\textbf {\bibinfo {volume}
  {19}},\ \bibinfo {pages} {35--50} (\bibinfo {year} {1961})}\BibitemShut
  {NoStop}%
\bibitem [{\citenamefont {Otsuji}\ \emph {et~al.}(2007)\citenamefont {Otsuji},
  \citenamefont {Ishihara}, \citenamefont {Co}, \citenamefont {Kaibuchi},
  \citenamefont {Mochizuki},\ and\ \citenamefont {Kuroda}}]{Otsuji.etal2007}%
  \BibitemOpen
  \bibfield  {author} {\bibinfo {author} {\bibfnamefont {Mikiya}\ \bibnamefont
  {Otsuji}}, \bibinfo {author} {\bibfnamefont {Shuji}\ \bibnamefont
  {Ishihara}}, \bibinfo {author} {\bibfnamefont {Carl}\ \bibnamefont {Co}},
  \bibinfo {author} {\bibfnamefont {Kozo}\ \bibnamefont {Kaibuchi}}, \bibinfo
  {author} {\bibfnamefont {Atsushi}\ \bibnamefont {Mochizuki}}, \ and\ \bibinfo
  {author} {\bibfnamefont {Shinya}\ \bibnamefont {Kuroda}},\ }\bibfield
  {title} {\enquote {\bibinfo {title} {A {{Mass Conserved
  Reaction}}\textendash{{Diffusion System Captures Properties}} of {{Cell
  Polarity}}},}\ }\href {\doibase 10.1371/journal.pcbi.0030108} {\bibfield
  {journal} {\bibinfo  {journal} {PLoS Computational Biology}\ }\textbf
  {\bibinfo {volume} {3}},\ \bibinfo {pages} {e108} (\bibinfo {year}
  {2007})}\BibitemShut {NoStop}%
\bibitem [{\citenamefont {Goryachev}\ and\ \citenamefont
  {Pokhilko}(2008)}]{Goryachev.Pokhilko2008}%
  \BibitemOpen
  \bibfield  {author} {\bibinfo {author} {\bibfnamefont {Andrew~B.}\
  \bibnamefont {Goryachev}}\ and\ \bibinfo {author} {\bibfnamefont
  {Alexandra~V.}\ \bibnamefont {Pokhilko}},\ }\bibfield  {title} {\enquote
  {\bibinfo {title} {Dynamics of {{Cdc42}} network embodies a {{Turing}}-type
  mechanism of yeast cell polarity},}\ }\href {\doibase
  10.1016/j.febslet.2008.03.029} {\bibfield  {journal} {\bibinfo  {journal}
  {FEBS Letters}\ }\textbf {\bibinfo {volume} {582}},\ \bibinfo {pages}
  {1437--1443} (\bibinfo {year} {2008})}\BibitemShut {NoStop}%
\bibitem [{\citenamefont {Altschuler}\ \emph {et~al.}(2008)\citenamefont
  {Altschuler}, \citenamefont {Angenent}, \citenamefont {Wang},\ and\
  \citenamefont {Wu}}]{Altschuler.etal2008}%
  \BibitemOpen
  \bibfield  {author} {\bibinfo {author} {\bibfnamefont {Steven~J.}\
  \bibnamefont {Altschuler}}, \bibinfo {author} {\bibfnamefont {Sigurd~B.}\
  \bibnamefont {Angenent}}, \bibinfo {author} {\bibfnamefont {Yanqin}\
  \bibnamefont {Wang}}, \ and\ \bibinfo {author} {\bibfnamefont {Lani~F.}\
  \bibnamefont {Wu}},\ }\bibfield  {title} {\enquote {\bibinfo {title} {On the
  spontaneous emergence of cell polarity},}\ }\href {\doibase
  10.1038/nature07119} {\bibfield  {journal} {\bibinfo  {journal} {Nature}\
  }\textbf {\bibinfo {volume} {454}},\ \bibinfo {pages} {886--889} (\bibinfo
  {year} {2008})}\BibitemShut {NoStop}%
\bibitem [{\citenamefont {Mori}\ \emph {et~al.}(2008)\citenamefont {Mori},
  \citenamefont {Jilkine},\ and\ \citenamefont
  {{Edelstein-Keshet}}}]{Mori.etal2008}%
  \BibitemOpen
  \bibfield  {author} {\bibinfo {author} {\bibfnamefont {Yoichiro}\
  \bibnamefont {Mori}}, \bibinfo {author} {\bibfnamefont {Alexandra}\
  \bibnamefont {Jilkine}}, \ and\ \bibinfo {author} {\bibfnamefont {Leah}\
  \bibnamefont {{Edelstein-Keshet}}},\ }\bibfield  {title} {\enquote {\bibinfo
  {title} {Wave-{{Pinning}} and {{Cell Polarity}} from a {{Bistable
  Reaction}}-{{Diffusion System}}},}\ }\href {\doibase
  10.1529/biophysj.107.120824} {\bibfield  {journal} {\bibinfo  {journal}
  {Biophysical Journal}\ }\textbf {\bibinfo {volume} {94}},\ \bibinfo {pages}
  {3684--3697} (\bibinfo {year} {2008})}\BibitemShut {NoStop}%
\bibitem [{\citenamefont {Jilkine}\ and\ \citenamefont
  {{Edelstein-Keshet}}(2011)}]{Jilkine.Edelstein-Keshet2011}%
  \BibitemOpen
  \bibfield  {author} {\bibinfo {author} {\bibfnamefont {Alexandra}\
  \bibnamefont {Jilkine}}\ and\ \bibinfo {author} {\bibfnamefont {Leah}\
  \bibnamefont {{Edelstein-Keshet}}},\ }\bibfield  {title} {\enquote {\bibinfo
  {title} {A {{Comparison}} of {{Mathematical Models}} for {{Polarization}} of
  {{Single Eukaryotic Cells}} in {{Response}} to {{Guided Cues}}},}\ }\href
  {\doibase 10.1371/journal.pcbi.1001121} {\bibfield  {journal} {\bibinfo
  {journal} {PLoS Computational Biology}\ }\textbf {\bibinfo {volume} {7}},\
  \bibinfo {pages} {e1001121} (\bibinfo {year} {2011})}\BibitemShut {NoStop}%
\bibitem [{\citenamefont {Trong}\ \emph {et~al.}(2014)\citenamefont {Trong},
  \citenamefont {Nicola}, \citenamefont {Goehring}, \citenamefont {Kumar},\
  and\ \citenamefont {Grill}}]{Trong.etal2014}%
  \BibitemOpen
  \bibfield  {author} {\bibinfo {author} {\bibfnamefont {Philipp~Khuc}\
  \bibnamefont {Trong}}, \bibinfo {author} {\bibfnamefont {Ernesto~M}\
  \bibnamefont {Nicola}}, \bibinfo {author} {\bibfnamefont {Nathan~W}\
  \bibnamefont {Goehring}}, \bibinfo {author} {\bibfnamefont {K~Vijay}\
  \bibnamefont {Kumar}}, \ and\ \bibinfo {author} {\bibfnamefont {Stephan~W}\
  \bibnamefont {Grill}},\ }\bibfield  {title} {\enquote {\bibinfo {title}
  {Parameter-space topology of models for cell polarity},}\ }\href {\doibase
  10.1088/1367-2630/16/6/065009} {\bibfield  {journal} {\bibinfo  {journal}
  {New Journal of Physics}\ }\textbf {\bibinfo {volume} {16}},\ \bibinfo
  {pages} {065009} (\bibinfo {year} {2014})}\BibitemShut {NoStop}%
\bibitem [{\citenamefont {Chiou}\ \emph {et~al.}(2018)\citenamefont {Chiou},
  \citenamefont {Ramirez}, \citenamefont {Elston}, \citenamefont {Witelski},
  \citenamefont {Schaeffer},\ and\ \citenamefont {Lew}}]{Chiou.etal2018}%
  \BibitemOpen
  \bibfield  {author} {\bibinfo {author} {\bibfnamefont {Jian-Geng}\
  \bibnamefont {Chiou}}, \bibinfo {author} {\bibfnamefont {Samuel~A.}\
  \bibnamefont {Ramirez}}, \bibinfo {author} {\bibfnamefont {Timothy~C.}\
  \bibnamefont {Elston}}, \bibinfo {author} {\bibfnamefont {Thomas~P.}\
  \bibnamefont {Witelski}}, \bibinfo {author} {\bibfnamefont {David~G.}\
  \bibnamefont {Schaeffer}}, \ and\ \bibinfo {author} {\bibfnamefont
  {Daniel~J.}\ \bibnamefont {Lew}},\ }\bibfield  {title} {\enquote {\bibinfo
  {title} {Principles that govern competition or co-existence in
  {{Rho}}-{{GTPase}} driven polarization},}\ }\href {\doibase
  10.1371/journal.pcbi.1006095} {\bibfield  {journal} {\bibinfo  {journal}
  {PLOS Computational Biology}\ }\textbf {\bibinfo {volume} {14}},\ \bibinfo
  {pages} {e1006095} (\bibinfo {year} {2018})}\BibitemShut {NoStop}%
\bibitem [{\citenamefont {Scheel}(2009)}]{Scheel2009}%
  \BibitemOpen
  \bibfield  {author} {\bibinfo {author} {\bibfnamefont {Arnd}\ \bibnamefont
  {Scheel}},\ }\bibfield  {title} {\enquote {\bibinfo {title} {Robustness of
  {{Liesegang}} patterns},}\ }\href {\doibase 10.1088/0951-7715/22/2/012}
  {\bibfield  {journal} {\bibinfo  {journal} {Nonlinearity}\ }\textbf {\bibinfo
  {volume} {22}},\ \bibinfo {pages} {457--483} (\bibinfo {year}
  {2009})}\BibitemShut {NoStop}%
\bibitem [{\citenamefont {Aranson}\ and\ \citenamefont
  {Tsimring}(2008)}]{Aranson.Tsimring2008}%
  \BibitemOpen
  \bibfield  {author} {\bibinfo {author} {\bibfnamefont {Igor}\ \bibnamefont
  {Aranson}}\ and\ \bibinfo {author} {\bibfnamefont {Lev}\ \bibnamefont
  {Tsimring}},\ }\href@noop {} {\emph {\bibinfo {title} {Granular
  {{Patterns}}}}}\ (\bibinfo  {publisher} {{Oxford University Press}},\
  \bibinfo {year} {2008})\BibitemShut {NoStop}%
\bibitem [{\citenamefont {Forte}\ \emph {et~al.}(2019)\citenamefont {Forte},
  \citenamefont {Caraglio}, \citenamefont {Marenduzzo},\ and\ \citenamefont
  {Orlandini}}]{Forte.etal2019}%
  \BibitemOpen
  \bibfield  {author} {\bibinfo {author} {\bibfnamefont {Giada}\ \bibnamefont
  {Forte}}, \bibinfo {author} {\bibfnamefont {Michele}\ \bibnamefont
  {Caraglio}}, \bibinfo {author} {\bibfnamefont {Davide}\ \bibnamefont
  {Marenduzzo}}, \ and\ \bibinfo {author} {\bibfnamefont {Enzo}\ \bibnamefont
  {Orlandini}},\ }\bibfield  {title} {\enquote {\bibinfo {title} {Plectoneme
  dynamics and statistics in braided polymers},}\ }\href {\doibase
  10.1103/PhysRevE.99.052503} {\bibfield  {journal} {\bibinfo  {journal}
  {Physical Review E}\ }\textbf {\bibinfo {volume} {99}},\ \bibinfo {pages}
  {052503} (\bibinfo {year} {2019})}\BibitemShut {NoStop}%
\bibitem [{\citenamefont {Ishihara}\ \emph {et~al.}(2007)\citenamefont
  {Ishihara}, \citenamefont {Otsuji},\ and\ \citenamefont
  {Mochizuki}}]{Ishihara.etal2007}%
  \BibitemOpen
  \bibfield  {author} {\bibinfo {author} {\bibfnamefont {Shuji}\ \bibnamefont
  {Ishihara}}, \bibinfo {author} {\bibfnamefont {Mikiya}\ \bibnamefont
  {Otsuji}}, \ and\ \bibinfo {author} {\bibfnamefont {Atsushi}\ \bibnamefont
  {Mochizuki}},\ }\bibfield  {title} {\enquote {\bibinfo {title} {Transient and
  steady state of mass-conserved reaction-diffusion systems},}\ }\href
  {\doibase 10.1103/PhysRevE.75.015203} {\bibfield  {journal} {\bibinfo
  {journal} {Physical Review E}\ }\textbf {\bibinfo {volume} {75}},\ \bibinfo
  {pages} {015203(R)} (\bibinfo {year} {2007})}\BibitemShut {NoStop}%
\bibitem [{\citenamefont {Morita}\ and\ \citenamefont
  {Ogawa}(2010)}]{Morita.Ogawa2010}%
  \BibitemOpen
  \bibfield  {author} {\bibinfo {author} {\bibfnamefont {Yoshihisa}\
  \bibnamefont {Morita}}\ and\ \bibinfo {author} {\bibfnamefont {Toshiyuki}\
  \bibnamefont {Ogawa}},\ }\bibfield  {title} {\enquote {\bibinfo {title}
  {Stability and bifurcation of nonconstant solutions to a reaction\textendash
  diffusion system with conservation of mass},}\ }\href {\doibase
  10.1088/0951-7715/23/6/007} {\bibfield  {journal} {\bibinfo  {journal}
  {Nonlinearity}\ }\textbf {\bibinfo {volume} {23}},\ \bibinfo {pages}
  {1387--1411} (\bibinfo {year} {2010})}\BibitemShut {NoStop}%
\bibitem [{\citenamefont {Brauns}\ \emph {et~al.}(2020)\citenamefont {Brauns},
  \citenamefont {Halatek},\ and\ \citenamefont {Frey}}]{Brauns.etal2020d}%
  \BibitemOpen
  \bibfield  {author} {\bibinfo {author} {\bibfnamefont {Fridtjof}\
  \bibnamefont {Brauns}}, \bibinfo {author} {\bibfnamefont {Jacob}\
  \bibnamefont {Halatek}}, \ and\ \bibinfo {author} {\bibfnamefont {Erwin}\
  \bibnamefont {Frey}},\ }\bibfield  {title} {\enquote {\bibinfo {title}
  {Phase-{{Space Geometry}} of {{Mass}}-{{Conserving Reaction}}-{{Diffusion
  Dynamics}}},}\ }\href {\doibase 10.1103/PhysRevX.10.041036} {\bibfield
  {journal} {\bibinfo  {journal} {Physical Review X}\ }\textbf {\bibinfo
  {volume} {10}},\ \bibinfo {pages} {041036} (\bibinfo {year}
  {2020})}\BibitemShut {NoStop}%
\bibitem [{\citenamefont {Cates}\ \emph {et~al.}(2010)\citenamefont {Cates},
  \citenamefont {Marenduzzo}, \citenamefont {Pagonabarraga},\ and\
  \citenamefont {Tailleur}}]{Cates.etal2010}%
  \BibitemOpen
  \bibfield  {author} {\bibinfo {author} {\bibfnamefont {M.~E.}\ \bibnamefont
  {Cates}}, \bibinfo {author} {\bibfnamefont {D.}~\bibnamefont {Marenduzzo}},
  \bibinfo {author} {\bibfnamefont {I.}~\bibnamefont {Pagonabarraga}}, \ and\
  \bibinfo {author} {\bibfnamefont {J.}~\bibnamefont {Tailleur}},\ }\bibfield
  {title} {\enquote {\bibinfo {title} {Arrested phase separation in reproducing
  bacteria creates a generic route to pattern formation},}\ }\href {\doibase
  10.1073/pnas.1001994107} {\bibfield  {journal} {\bibinfo  {journal}
  {Proceedings of the National Academy of Sciences}\ }\textbf {\bibinfo
  {volume} {107}},\ \bibinfo {pages} {11715--11720} (\bibinfo {year}
  {2010})}\BibitemShut {NoStop}%
\bibitem [{\citenamefont {Cates}\ and\ \citenamefont
  {Tailleur}(2015)}]{Cates.Tailleur2015}%
  \BibitemOpen
  \bibfield  {author} {\bibinfo {author} {\bibfnamefont {Michael~E.}\
  \bibnamefont {Cates}}\ and\ \bibinfo {author} {\bibfnamefont {Julien}\
  \bibnamefont {Tailleur}},\ }\bibfield  {title} {\enquote {\bibinfo {title}
  {Motility-{{Induced Phase Separation}}},}\ }\href {\doibase
  10.1146/annurev-conmatphys-031214-014710} {\bibfield  {journal} {\bibinfo
  {journal} {Annual Review of Condensed Matter Physics}\ }\textbf {\bibinfo
  {volume} {6}},\ \bibinfo {pages} {219--244} (\bibinfo {year}
  {2015})}\BibitemShut {NoStop}%
\bibitem [{\citenamefont {Liebchen}\ and\ \citenamefont
  {Levis}(2017)}]{Liebchen.Levis2017}%
  \BibitemOpen
  \bibfield  {author} {\bibinfo {author} {\bibfnamefont {Benno}\ \bibnamefont
  {Liebchen}}\ and\ \bibinfo {author} {\bibfnamefont {Demian}\ \bibnamefont
  {Levis}},\ }\bibfield  {title} {\enquote {\bibinfo {title} {Collective
  {{Behavior}} of {{Chiral Active Matter}}: {{Pattern Formation}} and
  {{Enhanced Flocking}}},}\ }\href {\doibase 10.1103/PhysRevLett.119.058002}
  {\bibfield  {journal} {\bibinfo  {journal} {Physical Review Letters}\
  }\textbf {\bibinfo {volume} {119}},\ \bibinfo {pages} {058002} (\bibinfo
  {year} {2017})}\BibitemShut {NoStop}%
\bibitem [{\citenamefont {Chat{\'e}}(2020)}]{Chate2020}%
  \BibitemOpen
  \bibfield  {author} {\bibinfo {author} {\bibfnamefont {Hugues}\ \bibnamefont
  {Chat{\'e}}},\ }\bibfield  {title} {\enquote {\bibinfo {title} {Dry
  {{Aligning Dilute Active Matter}}},}\ }\href {\doibase
  10.1146/annurev-conmatphys-031119-050752} {\bibfield  {journal} {\bibinfo
  {journal} {Annual Review of Condensed Matter Physics}\ }\textbf {\bibinfo
  {volume} {11}},\ \bibinfo {pages} {189--212} (\bibinfo {year}
  {2020})}\BibitemShut {NoStop}%
\bibitem [{Note1()}]{Note1}%
  \BibitemOpen
  \bibinfo {note} {We discuss the dispersion relation of Eq.~\protect \textup
  {\hbox {\mathsurround \z@ \protect \normalfont (\ignorespaces \ref
  {eq:2CRD}\unskip \@@italiccorr )}} linearized around a homogeneous steady
  state in Fig.~S1 in the SM.}\BibitemShut {Stop}%
\bibitem [{Note2()}]{Note2}%
  \BibitemOpen
  \bibinfo {note} {Our findings immediately generalize to systems with
  density-independent cross-diffusion, see SM Sec.~1.2.}\BibitemShut {Stop}%
\bibitem [{Note:SM()}]{Note:SM}%
  \BibitemOpen
  \bibinfo {note} {See Supplemental Material for Movies 1 and 2 as well as
  technical background information, which includes
  Refs.~\cite{Onsager1931,Hohenberg.Halperin1977,Bray2002,Alikakos.etal2004,Bressloff2020,Tateno.Ishihara2020,Rubinstein.Sternberg1992,Aranson.etal2002,Edelstein-Keshet.etal2013,Diegmiller.etal2018,Mikhailov1990,Zwicker.etal2017,Shin.Brangwynne2017,Gomes.Shorter2019}.}\BibitemShut
  {Stop}%
\bibitem [{Note3()}]{Note3}%
  \BibitemOpen
  \bibinfo {note} {Diffusive flux balance ensures that any stationary pattern
  can be dissected by inserting no-flux boundaries at its extrema.}\BibitemShut
  {Stop}%
\bibitem [{\citenamefont {Glasner}\ and\ \citenamefont
  {Witelski}(2003)}]{Glasner.Witelski2003}%
  \BibitemOpen
  \bibfield  {author} {\bibinfo {author} {\bibfnamefont {K.~B.}\ \bibnamefont
  {Glasner}}\ and\ \bibinfo {author} {\bibfnamefont {T.~P.}\ \bibnamefont
  {Witelski}},\ }\bibfield  {title} {\enquote {\bibinfo {title} {Coarsening
  dynamics of dewetting films},}\ }\href {\doibase 10.1103/PhysRevE.67.016302}
  {\bibfield  {journal} {\bibinfo  {journal} {Physical Review E}\ }\textbf
  {\bibinfo {volume} {67}},\ \bibinfo {pages} {016302} (\bibinfo {year}
  {2003})}\BibitemShut {NoStop}%
\bibitem [{\citenamefont {Pismen}\ and\ \citenamefont
  {Pomeau}(2004)}]{Pismen.Pomeau2004}%
  \BibitemOpen
  \bibfield  {author} {\bibinfo {author} {\bibfnamefont {Len~M.}\ \bibnamefont
  {Pismen}}\ and\ \bibinfo {author} {\bibfnamefont {Yves}\ \bibnamefont
  {Pomeau}},\ }\bibfield  {title} {\enquote {\bibinfo {title} {Mobility and
  interactions of weakly nonwetting droplets},}\ }\href {\doibase
  10.1063/1.1758911} {\bibfield  {journal} {\bibinfo  {journal} {Physics of
  Fluids}\ }\textbf {\bibinfo {volume} {16}},\ \bibinfo {pages} {2604--2612}
  (\bibinfo {year} {2004})}\BibitemShut {NoStop}%
\bibitem [{Note4()}]{Note4}%
  \BibitemOpen
  \bibinfo {note} {For mesa patterns in 1D, $\partial _M\eta _\protect \mathrm
  {stat}$ must be calculated for the high- and low-density plateaus separately
  (see SM Sec.~5).}\BibitemShut {Stop}%
\bibitem [{Note5()}]{Note5}%
  \BibitemOpen
  \bibinfo {note} {For a back of the envelope calculation we use density at the
  pattern inflection point $u_0$ and the interface width approximation $\ell
  _\protect \mathrm {int} \protect \tmspace +\thinmuskip {.1667em}{\approx
  }\protect \tmspace +\thinmuskip {.1667em} \pi /q_\protect \mathrm
  {max}|_{u_0, \eta _\protect \mathrm {stat}}$ \cite {Brauns.etal2020d}, and
  $\protect \mathaccentV {hat}05E{u} \approx 2 u_0$ to estimate the peak mass
  $M \protect \tmspace +\thinmuskip {.1667em}{=}\protect \tmspace +\thinmuskip
  {.1667em} \DOTSI \intop \ilimits@ \protect \mathrm {d}x [\protect
  \mathaccentV {tilde}07E{\rho }(x) \protect \tmspace +\thinmuskip
  {.1667em}{-}\protect \tmspace +\thinmuskip {.1667em} \rho _-] \protect
  \tmspace +\thinmuskip {.1667em}{\approx }\protect \tmspace +\thinmuskip
  {.1667em} \protect \mathaccentV {hat}05E{u} \cdot \ell _\protect \mathrm
  {int} \protect \tmspace +\thinmuskip {.1667em}{\approx }\protect \tmspace
  +\thinmuskip {.1667em} u_0 \cdot 2 \pi /q_\protect \mathrm {max}|_{\eta
  _\protect \mathrm {stat}} \protect \tmspace +\thinmuskip {.1667em}{\approx
  }\protect \tmspace +\thinmuskip {.1667em} 2\pi u_0 \protect \sqrt
  {D_u/f_u|_{\eta _\protect \mathrm {stat}}}$. For the example $f_\protect
  \mathrm {ex}$, one finds $f_u \protect \tmspace +\thinmuskip
  {.1667em}{\approx }\protect \tmspace +\thinmuskip {.1667em} \eta _\protect
  \mathrm {stat}$ and $u_0^{} \protect \tmspace +\thinmuskip {.1667em}{\approx
  }\protect \tmspace +\thinmuskip {.1667em} \eta _\protect \mathrm
  {stat}^{-1}$, and thus $M \protect \tmspace +\thinmuskip {.1667em}{\sim
  }\protect \tmspace +\thinmuskip {.1667em} \eta _\protect \mathrm
  {stat}^{-3/2} \protect \tmspace +\thinmuskip {.1667em}{=}\protect \tmspace
  +\thinmuskip {.1667em} \eta _\protect \mathrm {stat}^{-1/\alpha }$, for
  sufficiently large $M$ and $D_v \protect \tmspace +\thinmuskip {.1667em}{\gg
  }\protect \tmspace +\thinmuskip {.1667em} D_u$. Other reaction terms, e.g.\
  with different nonlinearities in the recruitment term, yields other exponents
  $\alpha $.}\BibitemShut {Stop}%
\bibitem [{Note6()}]{Note6}%
  \BibitemOpen
  \bibinfo {note} {If $u,v$ describe (shifted) concentrations ($u,v$ bounded
  from below) and $D_u/D_v$ is finite, mesas inevitably form at high densities
  (cf.\ Fig.~\ref {fig:peak-and-mesa-coarsening}a). Arbitrarily large peaks
  form if no third FBS-NC intersection point exists; as, for example, in
  ``Model II'' in Ref.~\cite {Otsuji.etal2007}. The peak-to-mesa transition has
  previously been observed for unstable thin films subject to gravity \cite
  {Gratton.Witelski2008}.}\BibitemShut {Stop}%
\bibitem [{\citenamefont {Gratton}\ and\ \citenamefont
  {Witelski}(2008)}]{Gratton.Witelski2008}%
  \BibitemOpen
  \bibfield  {author} {\bibinfo {author} {\bibfnamefont {M.~B.}\ \bibnamefont
  {Gratton}}\ and\ \bibinfo {author} {\bibfnamefont {T.~P.}\ \bibnamefont
  {Witelski}},\ }\bibfield  {title} {\enquote {\bibinfo {title} {Coarsening of
  unstable thin films subject to gravity},}\ }\href {\doibase
  10.1103/PhysRevE.77.016301} {\bibfield  {journal} {\bibinfo  {journal}
  {Physical Review E}\ }\textbf {\bibinfo {volume} {77}},\ \bibinfo {pages}
  {016301} (\bibinfo {year} {2008})}\BibitemShut {NoStop}%
\bibitem [{\citenamefont {Langer}(1971)}]{Langer1971}%
  \BibitemOpen
  \bibfield  {author} {\bibinfo {author} {\bibfnamefont {J.S}\ \bibnamefont
  {Langer}},\ }\bibfield  {title} {\enquote {\bibinfo {title} {Theory of
  spinodal decomposition in alloys},}\ }\href {\doibase
  10.1016/0003-4916(71)90162-X} {\bibfield  {journal} {\bibinfo  {journal}
  {Annals of Physics}\ }\textbf {\bibinfo {volume} {65}},\ \bibinfo {pages}
  {53--86} (\bibinfo {year} {1971})}\BibitemShut {NoStop}%
\bibitem [{\citenamefont {Kolokolnikov}\ \emph {et~al.}(2006)\citenamefont
  {Kolokolnikov}, \citenamefont {Erneux},\ and\ \citenamefont
  {Wei}}]{Kolokolnikov.etal2006}%
  \BibitemOpen
  \bibfield  {author} {\bibinfo {author} {\bibfnamefont {T.}~\bibnamefont
  {Kolokolnikov}}, \bibinfo {author} {\bibfnamefont {T.}~\bibnamefont
  {Erneux}}, \ and\ \bibinfo {author} {\bibfnamefont {J.}~\bibnamefont {Wei}},\
  }\bibfield  {title} {\enquote {\bibinfo {title} {Mesa-type patterns in the
  one-dimensional {{Brusselator}} and their stability},}\ }\href {\doibase
  10.1016/j.physd.2005.12.005} {\bibfield  {journal} {\bibinfo  {journal}
  {Physica D: Nonlinear Phenomena}\ }\textbf {\bibinfo {volume} {214}},\
  \bibinfo {pages} {63--77} (\bibinfo {year} {2006})}\BibitemShut {NoStop}%
\bibitem [{\citenamefont {Kolokolnikov}\ \emph {et~al.}(2007)\citenamefont
  {Kolokolnikov}, \citenamefont {Ward},\ and\ \citenamefont
  {Wei}}]{Kolokolnikov.etal2007}%
  \BibitemOpen
  \bibfield  {author} {\bibinfo {author} {\bibfnamefont {T.}~\bibnamefont
  {Kolokolnikov}}, \bibinfo {author} {\bibfnamefont {M.J.}\ \bibnamefont
  {Ward}}, \ and\ \bibinfo {author} {\bibfnamefont {J.}~\bibnamefont {Wei}},\
  }\bibfield  {title} {\enquote {\bibinfo {title} {Self-replication of mesa
  patterns in reaction\textendash diffusion systems},}\ }\href {\doibase
  10.1016/j.physd.2007.07.014} {\bibfield  {journal} {\bibinfo  {journal}
  {Physica D: Nonlinear Phenomena}\ }\textbf {\bibinfo {volume} {236}},\
  \bibinfo {pages} {104--122} (\bibinfo {year} {2007})}\BibitemShut {NoStop}%
\bibitem [{Note7()}]{Note7}%
  \BibitemOpen
  \bibinfo {note} {In steady state, net degradation in high-density regions
  ($\rho > \rho _\protect \mathrm {hss}$) and net production in low-density
  regions ($\rho < \rho _\protect \mathrm {hss}$) must balance.}\BibitemShut
  {Stop}%
\bibitem [{\citenamefont {Prigogine}\ and\ \citenamefont
  {Lefever}(1968)}]{Prigogine.Lefever1968}%
  \BibitemOpen
  \bibfield  {author} {\bibinfo {author} {\bibfnamefont {I.}~\bibnamefont
  {Prigogine}}\ and\ \bibinfo {author} {\bibfnamefont {R.}~\bibnamefont
  {Lefever}},\ }\bibfield  {title} {\enquote {\bibinfo {title} {Symmetry
  {{Breaking Instabilities}} in {{Dissipative Systems}}. {{II}}},}\ }\href
  {\doibase 10.1063/1.1668896} {\bibfield  {journal} {\bibinfo  {journal} {The
  Journal of Chemical Physics}\ }\textbf {\bibinfo {volume} {48}},\ \bibinfo
  {pages} {1695--1700} (\bibinfo {year} {1968})}\BibitemShut {NoStop}%
\bibitem [{\citenamefont {McKay}\ and\ \citenamefont
  {Kolokolnikov}(2012)}]{McKay.Kolokolnikov2012}%
  \BibitemOpen
  \bibfield  {author} {\bibinfo {author} {\bibfnamefont {Rebecca}\ \bibnamefont
  {McKay}}\ and\ \bibinfo {author} {\bibfnamefont {Theodore}\ \bibnamefont
  {Kolokolnikov}},\ }\bibfield  {title} {\enquote {\bibinfo {title} {Stability
  transitions and dynamics of mesa patterns near the shadow limit of
  reaction-diffusion systems in one space dimension},}\ }\href {\doibase
  10.3934/dcdsb.2012.17.191} {\bibfield  {journal} {\bibinfo  {journal}
  {Discrete \& Continuous Dynamical Systems - B}\ }\textbf {\bibinfo {volume}
  {17}},\ \bibinfo {pages} {191--220} (\bibinfo {year} {2012})}\BibitemShut
  {NoStop}%
\bibitem [{\citenamefont {Tjhung}\ \emph {et~al.}(2018)\citenamefont {Tjhung},
  \citenamefont {Nardini},\ and\ \citenamefont {Cates}}]{Tjhung.etal2018}%
  \BibitemOpen
  \bibfield  {author} {\bibinfo {author} {\bibfnamefont {Elsen}\ \bibnamefont
  {Tjhung}}, \bibinfo {author} {\bibfnamefont {Cesare}\ \bibnamefont
  {Nardini}}, \ and\ \bibinfo {author} {\bibfnamefont {Michael~E.}\
  \bibnamefont {Cates}},\ }\bibfield  {title} {\enquote {\bibinfo {title}
  {Cluster {{Phases}} and {{Bubbly Phase Separation}} in {{Active Fluids}}:
  {{Reversal}} of the {{Ostwald Process}}},}\ }\href {\doibase
  10.1103/PhysRevX.8.031080} {\bibfield  {journal} {\bibinfo  {journal}
  {Physical Review X}\ }\textbf {\bibinfo {volume} {8}},\ \bibinfo {pages}
  {031080} (\bibinfo {year} {2018})}\BibitemShut {NoStop}%
\bibitem [{\citenamefont {Vanag}\ and\ \citenamefont
  {Epstein}(2009)}]{Vanag.Epstein2009}%
  \BibitemOpen
  \bibfield  {author} {\bibinfo {author} {\bibfnamefont {Vladimir~K.}\
  \bibnamefont {Vanag}}\ and\ \bibinfo {author} {\bibfnamefont {Irving~R.}\
  \bibnamefont {Epstein}},\ }\bibfield  {title} {\enquote {\bibinfo {title}
  {Cross-diffusion and pattern formation in reaction\textendash diffusion
  systems},}\ }\href {\doibase 10.1039/B813825G} {\bibfield  {journal}
  {\bibinfo  {journal} {Phys. Chem. Chem. Phys.}\ }\textbf {\bibinfo {volume}
  {11}},\ \bibinfo {pages} {897--912} (\bibinfo {year} {2009})}\BibitemShut
  {NoStop}%
\bibitem [{\citenamefont {Rossi}\ \emph {et~al.}(2011)\citenamefont {Rossi},
  \citenamefont {Vanag},\ and\ \citenamefont {Epstein}}]{Rossi.etal2011}%
  \BibitemOpen
  \bibfield  {author} {\bibinfo {author} {\bibfnamefont {Federico}\
  \bibnamefont {Rossi}}, \bibinfo {author} {\bibfnamefont {Vladimir~K.}\
  \bibnamefont {Vanag}}, \ and\ \bibinfo {author} {\bibfnamefont {Irving~R.}\
  \bibnamefont {Epstein}},\ }\bibfield  {title} {\enquote {\bibinfo {title}
  {Pentanary {{Cross}}-{{Diffusion}} in {{Water}}-in-{{Oil Microemulsions
  Loaded}} with {{Two Components}} of the {{Belousov}}-{{Zhabotinsky
  Reaction}}},}\ }\href {\doibase 10.1002/chem.201002069} {\bibfield  {journal}
  {\bibinfo  {journal} {Chemistry - A European Journal}\ }\textbf {\bibinfo
  {volume} {17}},\ \bibinfo {pages} {2138--2145} (\bibinfo {year}
  {2011})}\BibitemShut {NoStop}%
\bibitem [{\citenamefont {Giri}\ \emph {et~al.}(2020)\citenamefont {Giri},
  \citenamefont {Pramod~Jain},\ and\ \citenamefont {Kar}}]{Giri.etal2020}%
  \BibitemOpen
  \bibfield  {author} {\bibinfo {author} {\bibfnamefont {Amitava}\ \bibnamefont
  {Giri}}, \bibinfo {author} {\bibfnamefont {Shreyans}\ \bibnamefont
  {Pramod~Jain}}, \ and\ \bibinfo {author} {\bibfnamefont {Sandip}\
  \bibnamefont {Kar}},\ }\bibfield  {title} {\enquote {\bibinfo {title}
  {Alteration in {{Cross Diffusivities Governs}} the {{Nature}} and
  {{Dynamics}} of {{Spatiotemporal Pattern Formation}}},}\ }\href {\doibase
  10.1002/cphc.202000142} {\bibfield  {journal} {\bibinfo  {journal}
  {ChemPhysChem}\ }\textbf {\bibinfo {volume} {21}},\ \bibinfo {pages}
  {1608--1616} (\bibinfo {year} {2020})}\BibitemShut {NoStop}%
\bibitem [{\citenamefont {Giunta}\ \emph {et~al.}(2020)\citenamefont {Giunta},
  \citenamefont {{Seyed-Allaei}},\ and\ \citenamefont
  {Gerland}}]{Giunta.etal2020}%
  \BibitemOpen
  \bibfield  {author} {\bibinfo {author} {\bibfnamefont {Giovanni}\
  \bibnamefont {Giunta}}, \bibinfo {author} {\bibfnamefont {Hamid}\
  \bibnamefont {{Seyed-Allaei}}}, \ and\ \bibinfo {author} {\bibfnamefont
  {Ulrich}\ \bibnamefont {Gerland}},\ }\bibfield  {title} {\enquote {\bibinfo
  {title} {Cross-diffusion induced patterns for a single-step enzymatic
  reaction},}\ }\href {\doibase 10.1038/s42005-020-00427-w} {\bibfield
  {journal} {\bibinfo  {journal} {Communications Physics}\ }\textbf {\bibinfo
  {volume} {3}},\ \bibinfo {pages} {167} (\bibinfo {year} {2020})}\BibitemShut
  {NoStop}%
\bibitem [{\citenamefont {Liu}\ \emph {et~al.}(2013)\citenamefont {Liu},
  \citenamefont {Doelman}, \citenamefont {Rottschafer}, \citenamefont {{de
  Jager}}, \citenamefont {Herman}, \citenamefont {Rietkerk},\ and\
  \citenamefont {{van de Koppel}}}]{Liu.etal2013}%
  \BibitemOpen
  \bibfield  {author} {\bibinfo {author} {\bibfnamefont {Q.-X.}\ \bibnamefont
  {Liu}}, \bibinfo {author} {\bibfnamefont {A.}~\bibnamefont {Doelman}},
  \bibinfo {author} {\bibfnamefont {V.}~\bibnamefont {Rottschafer}}, \bibinfo
  {author} {\bibfnamefont {M.}~\bibnamefont {{de Jager}}}, \bibinfo {author}
  {\bibfnamefont {P.~M.~J.}\ \bibnamefont {Herman}}, \bibinfo {author}
  {\bibfnamefont {M.}~\bibnamefont {Rietkerk}}, \ and\ \bibinfo {author}
  {\bibfnamefont {J.}~\bibnamefont {{van de Koppel}}},\ }\bibfield  {title}
  {\enquote {\bibinfo {title} {Phase separation explains a new class of
  self-organized spatial patterns in ecological systems},}\ }\href {\doibase
  10.1073/pnas.1222339110} {\bibfield  {journal} {\bibinfo  {journal}
  {Proceedings of the National Academy of Sciences}\ }\textbf {\bibinfo
  {volume} {110}},\ \bibinfo {pages} {11905--11910} (\bibinfo {year}
  {2013})}\BibitemShut {NoStop}%
\bibitem [{\citenamefont {Wittkowski}\ \emph {et~al.}(2014)\citenamefont
  {Wittkowski}, \citenamefont {Tiribocchi}, \citenamefont {Stenhammar},
  \citenamefont {Allen}, \citenamefont {Marenduzzo},\ and\ \citenamefont
  {Cates}}]{Wittkowski.etal2014}%
  \BibitemOpen
  \bibfield  {author} {\bibinfo {author} {\bibfnamefont {Raphael}\ \bibnamefont
  {Wittkowski}}, \bibinfo {author} {\bibfnamefont {Adriano}\ \bibnamefont
  {Tiribocchi}}, \bibinfo {author} {\bibfnamefont {Joakim}\ \bibnamefont
  {Stenhammar}}, \bibinfo {author} {\bibfnamefont {Rosalind~J.}\ \bibnamefont
  {Allen}}, \bibinfo {author} {\bibfnamefont {Davide}\ \bibnamefont
  {Marenduzzo}}, \ and\ \bibinfo {author} {\bibfnamefont {Michael~E.}\
  \bibnamefont {Cates}},\ }\bibfield  {title} {\enquote {\bibinfo {title}
  {Scalar {$\Phi$}4 field theory for active-particle phase separation},}\
  }\href {\doibase 10.1038/ncomms5351} {\bibfield  {journal} {\bibinfo
  {journal} {Nature Communications}\ }\textbf {\bibinfo {volume} {5}},\
  \bibinfo {pages} {4351} (\bibinfo {year} {2014})}\BibitemShut {NoStop}%
\bibitem [{\citenamefont {Gonnella}\ \emph {et~al.}(2015)\citenamefont
  {Gonnella}, \citenamefont {Marenduzzo}, \citenamefont {Suma},\ and\
  \citenamefont {Tiribocchi}}]{Gonnella.etal2015}%
  \BibitemOpen
  \bibfield  {author} {\bibinfo {author} {\bibfnamefont {Giuseppe}\
  \bibnamefont {Gonnella}}, \bibinfo {author} {\bibfnamefont {Davide}\
  \bibnamefont {Marenduzzo}}, \bibinfo {author} {\bibfnamefont {Antonio}\
  \bibnamefont {Suma}}, \ and\ \bibinfo {author} {\bibfnamefont {Adriano}\
  \bibnamefont {Tiribocchi}},\ }\bibfield  {title} {\enquote {\bibinfo {title}
  {Motility-induced phase separation and coarsening in active matter},}\ }\href
  {\doibase 10.1016/j.crhy.2015.05.001} {\bibfield  {journal} {\bibinfo
  {journal} {Comptes Rendus Physique}\ }\textbf {\bibinfo {volume} {16}},\
  \bibinfo {pages} {316--331} (\bibinfo {year} {2015})}\BibitemShut {NoStop}%
\bibitem [{\citenamefont {Sabrina}\ \emph {et~al.}(2015)\citenamefont
  {Sabrina}, \citenamefont {Spellings}, \citenamefont {Glotzer},\ and\
  \citenamefont {Bishop}}]{Sabrina.etal2015}%
  \BibitemOpen
  \bibfield  {author} {\bibinfo {author} {\bibfnamefont {Syeda}\ \bibnamefont
  {Sabrina}}, \bibinfo {author} {\bibfnamefont {Matthew}\ \bibnamefont
  {Spellings}}, \bibinfo {author} {\bibfnamefont {Sharon~C.}\ \bibnamefont
  {Glotzer}}, \ and\ \bibinfo {author} {\bibfnamefont {Kyle J.~M.}\
  \bibnamefont {Bishop}},\ }\bibfield  {title} {\enquote {\bibinfo {title}
  {Coarsening dynamics of binary liquids with active rotation},}\ }\href
  {\doibase 10.1039/C5SM01753J} {\bibfield  {journal} {\bibinfo  {journal}
  {Soft Matter}\ }\textbf {\bibinfo {volume} {11}},\ \bibinfo {pages}
  {8409--8416} (\bibinfo {year} {2015})}\BibitemShut {NoStop}%
\bibitem [{\citenamefont {Liu}\ \emph {et~al.}(2019)\citenamefont {Liu},
  \citenamefont {Patch}, \citenamefont {Bahar}, \citenamefont {Yllanes},
  \citenamefont {Welch}, \citenamefont {Marchetti}, \citenamefont
  {Thutupalli},\ and\ \citenamefont {Shaevitz}}]{Liu.etal2019}%
  \BibitemOpen
  \bibfield  {author} {\bibinfo {author} {\bibfnamefont {Guannan}\ \bibnamefont
  {Liu}}, \bibinfo {author} {\bibfnamefont {Adam}\ \bibnamefont {Patch}},
  \bibinfo {author} {\bibfnamefont {Fatmag{\"u}l}\ \bibnamefont {Bahar}},
  \bibinfo {author} {\bibfnamefont {David}\ \bibnamefont {Yllanes}}, \bibinfo
  {author} {\bibfnamefont {Roy~D.}\ \bibnamefont {Welch}}, \bibinfo {author}
  {\bibfnamefont {M.~Cristina}\ \bibnamefont {Marchetti}}, \bibinfo {author}
  {\bibfnamefont {Shashi}\ \bibnamefont {Thutupalli}}, \ and\ \bibinfo {author}
  {\bibfnamefont {Joshua~W.}\ \bibnamefont {Shaevitz}},\ }\bibfield  {title}
  {\enquote {\bibinfo {title} {Self-{{Driven Phase Transitions Drive}}
  {{{\emph{Myxococcus}}}}{\emph{ xanthus}} {{Fruiting Body Formation}}},}\
  }\href {\doibase 10.1103/PhysRevLett.122.248102} {\bibfield  {journal}
  {\bibinfo  {journal} {Physical Review Letters}\ }\textbf {\bibinfo {volume}
  {122}},\ \bibinfo {pages} {248102} (\bibinfo {year} {2019})}\BibitemShut
  {NoStop}%
\bibitem [{\citenamefont {Curatolo}\ \emph {et~al.}(2019)\citenamefont
  {Curatolo}, \citenamefont {Zhou}, \citenamefont {Zhao}, \citenamefont {Liu},
  \citenamefont {Daerr}, \citenamefont {Tailleur},\ and\ \citenamefont
  {Huang}}]{Curatolo.etal2019}%
  \BibitemOpen
  \bibfield  {author} {\bibinfo {author} {\bibfnamefont {A.~I.}\ \bibnamefont
  {Curatolo}}, \bibinfo {author} {\bibfnamefont {N.}~\bibnamefont {Zhou}},
  \bibinfo {author} {\bibfnamefont {Y.}~\bibnamefont {Zhao}}, \bibinfo {author}
  {\bibfnamefont {C.}~\bibnamefont {Liu}}, \bibinfo {author} {\bibfnamefont
  {A.}~\bibnamefont {Daerr}}, \bibinfo {author} {\bibfnamefont
  {J.}~\bibnamefont {Tailleur}}, \ and\ \bibinfo {author} {\bibfnamefont
  {J.}~\bibnamefont {Huang}},\ }\bibfield  {title} {\enquote {\bibinfo {title}
  {Engineering cooperative patterns in multi-species bacterial colonies},}\
  }\href {\doibase 10.1101/798827} {\bibfield  {journal} {\bibinfo  {journal}
  {bioRxiv}\ ,\ \bibinfo {pages} {doi:10.1101/798827}} (\bibinfo {year}
  {2019})}\BibitemShut {NoStop}%
\bibitem [{\citenamefont {Li}\ and\ \citenamefont
  {Cates}(2020)}]{Li.Cates2020}%
  \BibitemOpen
  \bibfield  {author} {\bibinfo {author} {\bibfnamefont {Yuting~I.}\
  \bibnamefont {Li}}\ and\ \bibinfo {author} {\bibfnamefont {Michael~E.}\
  \bibnamefont {Cates}},\ }\bibfield  {title} {\enquote {\bibinfo {title}
  {Non-equilibrium phase separation with reactions: {{A}} canonical model and
  its behaviour},}\ }\href@noop {} {\bibfield  {journal} {\bibinfo  {journal}
  {arXiv:2001.02563 [cond-mat]}\ } (\bibinfo {year} {2020})},\ \Eprint
  {http://arxiv.org/abs/2001.02563} {arXiv:2001.02563 [cond-mat]} \BibitemShut
  {NoStop}%
\bibitem [{\citenamefont {Onsager}(1931)}]{Onsager1931}%
  \BibitemOpen
  \bibfield  {author} {\bibinfo {author} {\bibfnamefont {Lars}\ \bibnamefont
  {Onsager}},\ }\bibfield  {title} {\enquote {\bibinfo {title} {Reciprocal
  {{Relations}} in {{Irreversible Processes}}. {{I}}.}}\ }\href {\doibase
  10.1103/PhysRev.37.405} {\bibfield  {journal} {\bibinfo  {journal} {Physical
  Review}\ }\textbf {\bibinfo {volume} {37}},\ \bibinfo {pages} {405--426}
  (\bibinfo {year} {1931})}\BibitemShut {NoStop}%
\bibitem [{\citenamefont {Hohenberg}\ and\ \citenamefont
  {Halperin}(1977)}]{Hohenberg.Halperin1977}%
  \BibitemOpen
  \bibfield  {author} {\bibinfo {author} {\bibfnamefont {P.~C.}\ \bibnamefont
  {Hohenberg}}\ and\ \bibinfo {author} {\bibfnamefont {B.~I.}\ \bibnamefont
  {Halperin}},\ }\bibfield  {title} {\enquote {\bibinfo {title} {Theory of
  dynamic critical phenomena},}\ }\href {\doibase 10.1103/RevModPhys.49.435}
  {\bibfield  {journal} {\bibinfo  {journal} {Reviews of Modern Physics}\
  }\textbf {\bibinfo {volume} {49}},\ \bibinfo {pages} {435--479} (\bibinfo
  {year} {1977})}\BibitemShut {NoStop}%
\bibitem [{\citenamefont {Bray}(2002)}]{Bray2002}%
  \BibitemOpen
  \bibfield  {author} {\bibinfo {author} {\bibfnamefont {A.~J.}\ \bibnamefont
  {Bray}},\ }\bibfield  {title} {\enquote {\bibinfo {title} {Theory of
  phase-ordering kinetics},}\ }\href {\doibase 10.1080/00018730110117433}
  {\bibfield  {journal} {\bibinfo  {journal} {Advances in Physics}\ }\textbf
  {\bibinfo {volume} {51}},\ \bibinfo {pages} {481--587} (\bibinfo {year}
  {2002})}\BibitemShut {NoStop}%
\bibitem [{\citenamefont {Alikakos}\ \emph {et~al.}(2004)\citenamefont
  {Alikakos}, \citenamefont {Fusco},\ and\ \citenamefont
  {Karali}}]{Alikakos.etal2004}%
  \BibitemOpen
  \bibfield  {author} {\bibinfo {author} {\bibfnamefont {Nicholas~D}\
  \bibnamefont {Alikakos}}, \bibinfo {author} {\bibfnamefont {Giorgio}\
  \bibnamefont {Fusco}}, \ and\ \bibinfo {author} {\bibfnamefont {Georgia}\
  \bibnamefont {Karali}},\ }\bibfield  {title} {\enquote {\bibinfo {title}
  {Ostwald ripening in two dimensions\textemdash the rigorous derivation of the
  equations from the {{Mullins}}\textendash{{Sekerka}} dynamics},}\ }\href
  {\doibase 10.1016/j.jde.2004.05.008} {\bibfield  {journal} {\bibinfo
  {journal} {Journal of Differential Equations}\ }\textbf {\bibinfo {volume}
  {205}},\ \bibinfo {pages} {1--49} (\bibinfo {year} {2004})}\BibitemShut
  {NoStop}%
\bibitem [{\citenamefont {Bressloff}(2020)}]{Bressloff2020}%
  \BibitemOpen
  \bibfield  {author} {\bibinfo {author} {\bibfnamefont {Paul~C}\ \bibnamefont
  {Bressloff}},\ }\bibfield  {title} {\enquote {\bibinfo {title}
  {Two-dimensional droplet ripening in a concentration gradient},}\ }\href
  {\doibase 10.1088/1751-8121/aba39a} {\bibfield  {journal} {\bibinfo
  {journal} {Journal of Physics A: Mathematical and Theoretical}\ }\textbf
  {\bibinfo {volume} {53}},\ \bibinfo {pages} {365002} (\bibinfo {year}
  {2020})}\BibitemShut {NoStop}%
\bibitem [{\citenamefont {Tateno}\ and\ \citenamefont
  {Ishihara}(2020)}]{Tateno.Ishihara2020}%
  \BibitemOpen
  \bibfield  {author} {\bibinfo {author} {\bibfnamefont {Michio}\ \bibnamefont
  {Tateno}}\ and\ \bibinfo {author} {\bibfnamefont {Shuji}\ \bibnamefont
  {Ishihara}},\ }\bibfield  {title} {\enquote {\bibinfo {title}
  {Surface-tension-driven coarsening in mass-conserved reaction-diffusion
  systems},}\ }\href@noop {} {\bibfield  {journal} {\bibinfo  {journal}
  {arXiv:2010.03900 [cond-mat, physics:physics]}\ } (\bibinfo {year} {2020})},\
  \Eprint {http://arxiv.org/abs/2010.03900} {arXiv:2010.03900 [cond-mat,
  physics:physics]} \BibitemShut {NoStop}%
\bibitem [{\citenamefont {Rubinstein}\ and\ \citenamefont
  {Sternberg}(1992)}]{Rubinstein.Sternberg1992}%
  \BibitemOpen
  \bibfield  {author} {\bibinfo {author} {\bibfnamefont {Jacob}\ \bibnamefont
  {Rubinstein}}\ and\ \bibinfo {author} {\bibfnamefont {Peter}\ \bibnamefont
  {Sternberg}},\ }\bibfield  {title} {\enquote {\bibinfo {title} {Nonlocal
  reaction\textemdash diffusion equations and nucleation},}\ }\href {\doibase
  10.1093/imamat/48.3.249} {\bibfield  {journal} {\bibinfo  {journal} {IMA
  Journal of Applied Mathematics}\ }\textbf {\bibinfo {volume} {48}},\ \bibinfo
  {pages} {249--264} (\bibinfo {year} {1992})}\BibitemShut {NoStop}%
\bibitem [{\citenamefont {Aranson}\ \emph {et~al.}(2002)\citenamefont
  {Aranson}, \citenamefont {Meerson}, \citenamefont {Sasorov},\ and\
  \citenamefont {Vinokur}}]{Aranson.etal2002}%
  \BibitemOpen
  \bibfield  {author} {\bibinfo {author} {\bibfnamefont {I.~S.}\ \bibnamefont
  {Aranson}}, \bibinfo {author} {\bibfnamefont {B.}~\bibnamefont {Meerson}},
  \bibinfo {author} {\bibfnamefont {P.~V.}\ \bibnamefont {Sasorov}}, \ and\
  \bibinfo {author} {\bibfnamefont {V.~M.}\ \bibnamefont {Vinokur}},\
  }\bibfield  {title} {\enquote {\bibinfo {title} {Phase {{Separation}} and
  {{Coarsening}} in {{Electrostatically Driven Granular Media}}},}\ }\href
  {\doibase 10.1103/PhysRevLett.88.204301} {\bibfield  {journal} {\bibinfo
  {journal} {Physical Review Letters}\ }\textbf {\bibinfo {volume} {88}},\
  \bibinfo {pages} {204301} (\bibinfo {year} {2002})}\BibitemShut {NoStop}%
\bibitem [{\citenamefont {{Edelstein-Keshet}}\ \emph
  {et~al.}(2013)\citenamefont {{Edelstein-Keshet}}, \citenamefont {Holmes},
  \citenamefont {Zajac},\ and\ \citenamefont
  {Dutot}}]{Edelstein-Keshet.etal2013}%
  \BibitemOpen
  \bibfield  {author} {\bibinfo {author} {\bibfnamefont {Leah}\ \bibnamefont
  {{Edelstein-Keshet}}}, \bibinfo {author} {\bibfnamefont {William~R.}\
  \bibnamefont {Holmes}}, \bibinfo {author} {\bibfnamefont {Mark}\ \bibnamefont
  {Zajac}}, \ and\ \bibinfo {author} {\bibfnamefont {Meghan}\ \bibnamefont
  {Dutot}},\ }\bibfield  {title} {\enquote {\bibinfo {title} {From simple to
  detailed models for cell polarization},}\ }\href {\doibase
  10.1098/rstb.2013.0003} {\bibfield  {journal} {\bibinfo  {journal}
  {Philosophical Transactions of the Royal Society B: Biological Sciences}\
  }\textbf {\bibinfo {volume} {368}},\ \bibinfo {pages} {20130003} (\bibinfo
  {year} {2013})}\BibitemShut {NoStop}%
\bibitem [{\citenamefont {Diegmiller}\ \emph {et~al.}(2018)\citenamefont
  {Diegmiller}, \citenamefont {Montanelli}, \citenamefont {Muratov},\ and\
  \citenamefont {Shvartsman}}]{Diegmiller.etal2018}%
  \BibitemOpen
  \bibfield  {author} {\bibinfo {author} {\bibfnamefont {Rocky}\ \bibnamefont
  {Diegmiller}}, \bibinfo {author} {\bibfnamefont {Hadrien}\ \bibnamefont
  {Montanelli}}, \bibinfo {author} {\bibfnamefont {Cyrill~B.}\ \bibnamefont
  {Muratov}}, \ and\ \bibinfo {author} {\bibfnamefont {Stanislav~Y.}\
  \bibnamefont {Shvartsman}},\ }\bibfield  {title} {\enquote {\bibinfo {title}
  {Spherical {{Caps}} in {{Cell Polarization}}},}\ }\href {\doibase
  10.1016/j.bpj.2018.05.033} {\bibfield  {journal} {\bibinfo  {journal}
  {Biophysical Journal}\ }\textbf {\bibinfo {volume} {115}},\ \bibinfo {pages}
  {26--30} (\bibinfo {year} {2018})}\BibitemShut {NoStop}%
\bibitem [{\citenamefont {Mikhailov}(1990)}]{Mikhailov1990}%
  \BibitemOpen
  \bibfield  {author} {\bibinfo {author} {\bibfnamefont {Alexander~S}\
  \bibnamefont {Mikhailov}},\ }\href@noop {} {\emph {\bibinfo {title}
  {Foundations of {{Synergetics I}}: {{Distributed Active Systems}}}}}\
  (\bibinfo  {publisher} {{Springer Berlin Heidelberg}},\ \bibinfo {address}
  {{Berlin, Heidelberg}},\ \bibinfo {year} {1990})\BibitemShut {NoStop}%
\bibitem [{\citenamefont {Zwicker}\ \emph {et~al.}(2017)\citenamefont
  {Zwicker}, \citenamefont {Seyboldt}, \citenamefont {Weber}, \citenamefont
  {Hyman},\ and\ \citenamefont {J{\"u}licher}}]{Zwicker.etal2017}%
  \BibitemOpen
  \bibfield  {author} {\bibinfo {author} {\bibfnamefont {David}\ \bibnamefont
  {Zwicker}}, \bibinfo {author} {\bibfnamefont {Rabea}\ \bibnamefont
  {Seyboldt}}, \bibinfo {author} {\bibfnamefont {Christoph~A.}\ \bibnamefont
  {Weber}}, \bibinfo {author} {\bibfnamefont {Anthony~A.}\ \bibnamefont
  {Hyman}}, \ and\ \bibinfo {author} {\bibfnamefont {Frank}\ \bibnamefont
  {J{\"u}licher}},\ }\bibfield  {title} {\enquote {\bibinfo {title} {Growth and
  division of active droplets provides a model for protocells},}\ }\href
  {\doibase 10.1038/nphys3984} {\bibfield  {journal} {\bibinfo  {journal}
  {Nature Physics}\ }\textbf {\bibinfo {volume} {13}},\ \bibinfo {pages}
  {408--413} (\bibinfo {year} {2017})}\BibitemShut {NoStop}%
\bibitem [{\citenamefont {Shin}\ and\ \citenamefont
  {Brangwynne}(2017)}]{Shin.Brangwynne2017}%
  \BibitemOpen
  \bibfield  {author} {\bibinfo {author} {\bibfnamefont {Yongdae}\ \bibnamefont
  {Shin}}\ and\ \bibinfo {author} {\bibfnamefont {Clifford~P.}\ \bibnamefont
  {Brangwynne}},\ }\bibfield  {title} {\enquote {\bibinfo {title} {Liquid phase
  condensation in cell physiology and disease},}\ }\href {\doibase
  10.1126/science.aaf4382} {\bibfield  {journal} {\bibinfo  {journal}
  {Science}\ }\textbf {\bibinfo {volume} {357}},\ \bibinfo {pages} {eaaf4382}
  (\bibinfo {year} {2017})}\BibitemShut {NoStop}%
\bibitem [{\citenamefont {Gomes}\ and\ \citenamefont
  {Shorter}(2019)}]{Gomes.Shorter2019}%
  \BibitemOpen
  \bibfield  {author} {\bibinfo {author} {\bibfnamefont {Edward}\ \bibnamefont
  {Gomes}}\ and\ \bibinfo {author} {\bibfnamefont {James}\ \bibnamefont
  {Shorter}},\ }\bibfield  {title} {\enquote {\bibinfo {title} {The molecular
  language of membraneless organelles},}\ }\href {\doibase
  10.1074/jbc.TM118.001192} {\bibfield  {journal} {\bibinfo  {journal} {Journal
  of Biological Chemistry}\ }\textbf {\bibinfo {volume} {294}},\ \bibinfo
  {pages} {7115--7127} (\bibinfo {year} {2019})}\BibitemShut {NoStop}%
\end{thebibliography}
%

\end{document}